\documentclass[pra, onecolumn,superscriptaddress,nofootinbib,longbibliography]{revtex4-1}
\usepackage[a4paper, left=1in, right=1in, top=1in, bottom=1in]{geometry}

\usepackage{graphicx}
\usepackage{epstopdf}
\usepackage[english]{babel}
\usepackage{amsmath}
\usepackage{amssymb}
\usepackage{mathtools}
\usepackage{times}
\usepackage{appendix}
\usepackage{bm}
\usepackage{float}
\usepackage{color}
\usepackage{longtable}
\usepackage{url}
\usepackage[usenames,dvipsnames]{xcolor}
\usepackage[colorlinks=true,linkcolor=Blue,urlcolor=Blue,citecolor=Blue]{hyperref}
\usepackage{amsthm}
\usepackage[dvipsnames]{xcolor}
\usepackage{lineno}

\makeatletter

\newtheorem{theorem}{Theorem}

\newtheorem{definition}{Definition}

\graphicspath{{./Figures}}

\newcommand{\mdf}[1]{{\color{black} #1}}

\begin{document}
\title{Quantum circuit complexity and unsupervised machine learning of topological order}
\date{\today}

\affiliation{Department of Physics, University of Michigan, Ann Arbor, Michigan
48109-1040, USA}
\affiliation{Center for Quantum Computing, RIKEN, Wako-shi, Saitama 351-0198, Japan}
\affiliation{Department of Physics, Zhejiang Sci-Tech University, Hangzhou 310018, China}

\author{Yanming Che}
\email{yanmingche01@gmail.com}
\affiliation{Department of Physics, University of Michigan, Ann Arbor, Michigan
48109-1040, USA}
\affiliation{Center for Quantum Computing, RIKEN, Wako-shi, Saitama 351-0198, Japan}

\author{Clemens Gneiting}
\email{clemens.gneiting@riken.jp}
\affiliation{Center for Quantum Computing, RIKEN, Wako-shi, Saitama 351-0198, Japan}

\author{Xiaoguang Wang}
\email{xgwang@zstu.edu.cn}
\affiliation{Department of Physics, Zhejiang Sci-Tech University, Hangzhou 310018, China}

\author{Franco Nori}
\email{fnori@riken.jp}
\affiliation{Center for Quantum Computing, RIKEN, Wako-shi, Saitama 351-0198, Japan}
\affiliation{Department of Physics, University of Michigan, Ann Arbor, Michigan
48109-1040, USA}

\begin{abstract}
Inspired by the close relationship between Kolmogorov complexity and unsupervised machine learning, we explore quantum {\it circuit} complexity, an important concept in quantum computation and quantum information science, as a pivot to understand and to build interpretable and efficient unsupervised machine learning for topological order in quantum many-body systems. \mdf{We argue that Nielsen's quantum circuit complexity represents an intrinsic topological distance or similarity measure between topological quantum many-body phases of matter, and as such plays a central role in {\it interpretable} manifold learning of topological order. }
To span a bridge from conceptual power to practical applicability, we present two theorems that connect Nielsen's quantum circuit complexity for the quantum path planning between two {\it arbitrary} quantum many-body states with \mdf{ quantum Fisher complexity (Bures distance) } and entanglement generation, respectively. Leveraging these connections, fidelity-based and entanglement-based similarity measures or kernels, which are more  practical for implementation, are formulated. Using the two proposed \mdf{distance measures, unsupervised manifold learning} of quantum phases of the bond-alternating XXZ spin chain, the ground state of Kitaev's toric code and random product states, is conducted, demonstrating their superior performance. \mdf{Moreover, we find that the entanglement-based approach, which captures the long-range structure of quantum entanglement of topological orders, is more robust to local Haar random noises.} Relations with classical shadow tomography and shadow kernel learning are also discussed, where the latter can be naturally understood from our approach. 
Our results establish connections between key concepts and tools of quantum circuit computation, quantum complexity, \mdf{quantum metrology,} and machine learning of topological quantum order.
\end{abstract}

\maketitle

\section{Introduction}
\label{sec:I}
The discovery of topological phases of matter has opened a new chapter in modern physics, ranging from symmetry-protected band topologies with short-range quantum entanglement such as topological insulators and superconductors~\cite{KaneRMP2010,QiRMP2010}, to topological quantum orders~\cite{ZengPRB2015,ZengArxiv2015,KitaevPRL2006,LevinPRL2006,KitaevAnnPhys2003,SemeghiniScience2021,SatzingerScience2021,GoogleQuantumAINature2024} featured by distributed non-local quantum entanglement such as in Kitaev's toric code~\cite{KitaevAnnPhys2003,SatzingerScience2021}. Defying Landau's paradigmatic approach, where phases and phase transitions are identified through local order parameters (that is, linear functionals of the density matrix of the phase and a local observable), topological phases encode the phase information in global --topological-- system properties. This defining difference renders the detection and measurement of topological order challenging for both generic simulation and experimental demonstration.

Recent progress has shown the potential of machine learning for classifying topological  phases of matter~\cite{CarleoTroyerScience2017,CarrasquillaNatPhys2017,NieuwenburgNatPhys2017,ZhangPRL2017Quantum,YeHuaPRL2018,HuiZhaiPRL2018,HuiZhaiPRB2018,CarleoRMPML,ming2019quantum,RemNatPhys2019,LianPRL2019,Rodriguez-NievaNatPhys2019,ChePRB2020,ScheurerPRL2020,LongPRL2020,Greplova2020,BalabanovPRR2020,DengPRL2021,Mendes-SantosPRX2021,YuNpjQI2022,LeykamTDA2023,DanielNano2024,YuFR2024,DasSarma2019}. In particular, the \textit{supervised} approach to quantum many-body physics problems, which can resort to well established theories of learnability and generalizability~\cite{Valiant1984,FoundationsMLBook,NTK2018}, has resulted in the formulation of rigorous guarantees~\cite{HuangScience2022Manybody,LewisNC2024,ChePRR2024,RouzeNC2024,ChoNC2024,ZhaoPRXQ2024,OnoratiArxiv2024,DuArxiv2024,WannerArxiv2024,arxiv:2312.17019}. However, in many practical scenarios, where a priori knowledge of the different phases of matter (and accordingly their labels) does not exist, only \textit{unsupervised} machine learning can provide access to the desired information. Despite the practical relevance of unsupervised learning, a rigorous theory, tailored to topological quantum order and offering both interpretability and generalizability, has not yet been fully explored.

\mdf{One prevalent approach to unsupervised learning is manifold learning~\cite{TenenbaumScience2000,RoweisScience2000,Maaten2008t-SNE}, where the data is assumed to be distributed on a manifold characterized by an intrinsic distance metric. It is followed by a nonlinear mapping of the data manifold into a low-dimensional Euclidean space while preserving the original similarities and structures. Among many manifold-learning algorithms, diffusion map~\cite{coifman2005geometric,nadler2006diffusion}, t-SNE~\cite{Maaten2008t-SNE}, and kernel principal component analysis (PCA)~\cite{Scholkopf1998Nonlinear,Mika1999Kernel} require a kernel function constructed from the intrinsic distance. The kernel can be used either to build a probabilistic graph representation of the data manifold (e.g., in the diffusion map or the t-SNE), or for PCA in the nonlinear feature space (e.g., in the kernel PCA). Alternatively, the intrinsic distance metric itself can be applied directly to non-kernel manifold learning, such as the  Isomap~\cite{TenenbaumScience2000} and the metric-multidimensional scaling (metric-MDS)~\cite{DemainePMLR2021,SklearnMDS}. } 
In the context of topological quantum order, the basic task then is to formulate a \mdf{distance metric} that encodes the nonlinear and non-Euclidean features of topological quantum states and represents topological similarities or distances \mdf{\textit{by exploiting the primitive notion of topological equivalence}.} Topologically equivalent quantum states, by definition, can be continuously transformed into each other via a smooth unitary transformation generated by a local Hermitian operator with bounded operator norm~\cite{ZengArxiv2015}. In this spirit, a similarity measure based on a path-finding algorithm has been proposed in~\cite{ScheurerPRL2020} and has shown good performance in clustering symmetry-protected band topologies, which are characterized by short-range quantum entanglement. Other topological similarity measures, that focus on the closing of the spectral gap at topological transitions, have also been proposed and demonstrated~\cite{ChePRB2020,LongPRL2023}. 
	
Despite the elegance of these methods in identifying band topologies, the increasing complexity and quantum entanglement in strongly interacting quantum many-body systems may undermine the efficiency and performance of these algorithms. For instance, the path-finding algorithm, when applied to interacting lattice models in order to identify a smooth unitary path generated by a local operator while keeping the parent Hamiltonian gapped, can be principally inconclusive due to the undecidability of the spectral gap for generic quantum many-body Hamiltonians~\cite{CubittNature2015}.
	
\begin{figure*}[t]
\centerline{\includegraphics[height=4in,width=6in,clip]{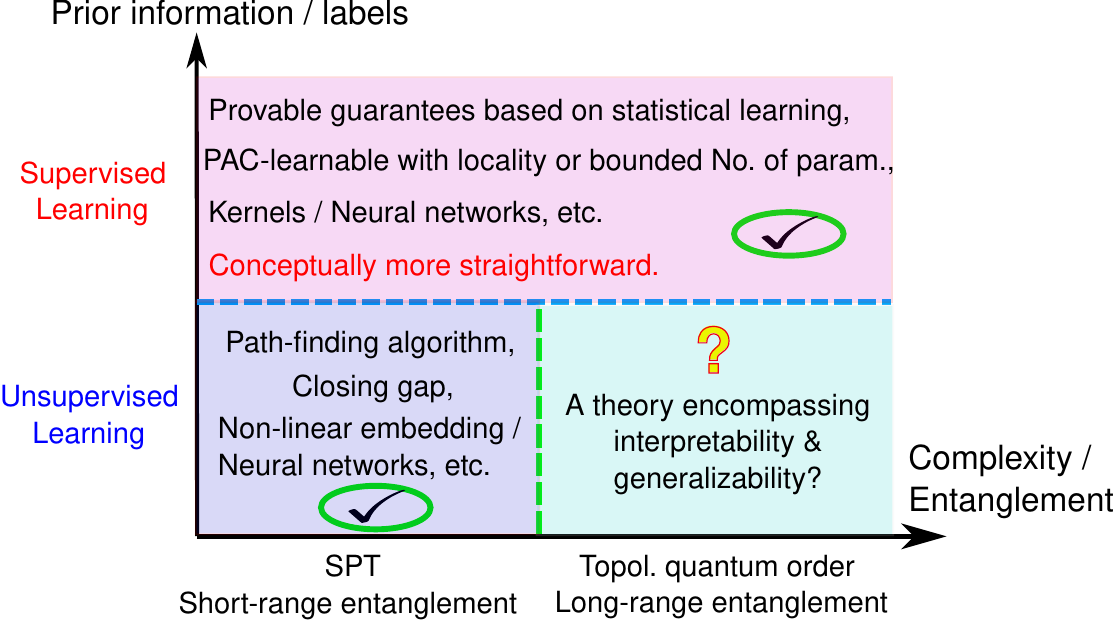}}
\caption{\textbf{Landscape of machine learning topological phases of matter.} Supervised learning requires prior information or labels, whose efficiency guarantees have been extensively investigated~\cite{HuangScience2022Manybody,LewisNC2024,ChePRR2024,RouzeNC2024,ChoNC2024,ZhaoPRXQ2024,OnoratiArxiv2024,DuArxiv2024,WannerArxiv2024,arxiv:2312.17019}. Under the constraint of geometric locality~\cite{LewisNC2024,RouzeNC2024} or with a bounded number (No.) of parameters (param.)~\cite{ChePRR2024}, provably efficient machine learning of quantum many-body systems has been established, e.g., in the framework of probably approximately correct (PAC)-learnable~\cite{Valiant1984,ChePRR2024}. Supervised learning is conceptually more straightforward due to the existence of well-established statistical learning theories. In contrast, unsupervised learning, whose theory is more complicated and less addressed, becomes practically relevant when there is no access to prior labels. 
The symmetry-protected topological (SPT) order features short-range entanglement, while the topological quantum order exhibits long-range entanglement. Unsupervised learning with the path-finding algorithm~\cite{ScheurerPRL2020} and kernels focusing on closing gaps at topological transitions~\cite{ChePRB2020,LongPRL2023}, respectively, have shown good performance and interpretability for SPT orders or band topologies. However, the direct generalization of these methods to more complex quantum systems with increased complexity and quantum entanglement faces challenges, due to the intrinsic hardness of identifying (or even the undecidability~\cite{CubittNature2015} of) the spectral gap for generic quantum many-body Hamiltonians (See more elaborations and references in the introduction of the main text). Note that reinforcement learning for quantum systems are not included in this landscape.
}
\label{fig:Schematic}
\end{figure*}

\mdf{The inherent problems of previous manifold-learning approaches call for a new similarity ansatz,} an ansatz which is able to capture the topological distance and entanglement patterns at different scales while at the same time remaining interpretable and efficiently computable (see Fig.~\ref{fig:Schematic} for a schematic of the current state regarding the application of machine learning to the classification of topological phases of matter). To this end, we here adopt the alternative but equivalent perspective that two topologically equivalent multi-qubit states can be transformed into each other via a \textit{constant-depth} quantum circuit composed of geometrically local quantum gates~\cite{ZengArxiv2015,ZengPRB2015}, i.e., the smooth unitary transformation that connects them can be decomposed into a shallow quantum circuit. This implies that it is algorithmically trivial or cheap to generate one quantum state given the access to another in the same topological class by fully exploiting the shared structure and entanglement patterns between them.

The correspondence between topological equivalence of quantum states and shallow quantum circuit transformations indicates that, a complexity measure may be related to the topological distance. This is similar to the situation with Kolmogorov complexity, which represents an informational distance between two strings or objects~\cite{AlgInfoDist}, and may offer a basis for theoretically optimal solution to unsupervised machine learning~\cite{IlyaNote} (See more details in the following section). We then assert that, the (conditional) generation and prediction of quantum states with the minimal quantum circuit cost may offer a theoretically optimal solution to the unsupervised learning of topological order. The complexity of the generating circuit can be interpreted as a quantum informational distance measure that captures the entanglement change at different scales. Specifically, we propose to use Nielsen's quantum circuit complexity (QCC) of the unitary path connecting two quantum states as a suitable distance measure that can be used for the unsupervised manifold learning of topological order. While the QCC is already a central concept in quantum information theory~\cite{YaoAFCS1993,BrandPRXQ2021,HaferkampNatPhys2022,LiuNatPhys2024,LiuPRR2020}, quantum field theory~\cite{ChapmanPRL2018,HacklJHEP2018,KhanPRD2018} and quantum gravity~\cite{SusskindArxiv2014,StanfordPRD2014,BrownPRL2016,BrownPRD2016,BrownNature2023}, its connection with, and relevance for, the unsupervised machine learning of topological phases of matter has remained mostly unexplored. 

In this work, we use the QCC as a pivotal tool to understand and construct unsupervised machine learning of topological order for quantum many-body systems, in an interpretable and efficient manner. Although the exact QCC is generically intractable, we identify various quantities which are upper bounded by the QCC while preserving useful information and remaining implementable, and which thus can be efficiently utilized for classifying topological quantum many-body phases. Specifically, we establish entanglement-based and fidelity-based similarities / kernels, respectively. Numerical experiments with a bond-alternating XXZ qubit chain and Kitaev's toric code model, respectively, demonstrate the excellent performance of these methods. 

\section{Results} 

\subsection{Quantum path planning, topological order and quantum circuit complexity}

As mentioned above, unsupervised machine learning of band topology has been extensively investigated. Band topologies, or the more general symmetry-protected topological (SPT) orders, such as those exhibited by topological insulators, are characterized by the absence of long-range quantum entanglement. This implies that, if the constraint of preserving the symmetry is dropped, SPT quantum states can be transformed smoothly into product states. In contrast, topological quantum order refers to quantum states with long-range quantum entanglement~\cite{ZengArxiv2015}; for instance, the ground state of Kitaev's toric code exhibits topological quantum order that manifests in non-zero topological entanglement entropy~\cite{KitaevPRL2006,LevinPRL2006} and global logical operations~\cite{KitaevAnnPhys2003}, while SPT phases are confined to short-range entanglement. Thus, the task of identifying topological quantum order with unsupervised learning presupposes to construct similarity measures with feature maps that capture quantum entanglement at a variety of scales. For this purpose, we start by defining the quantum path planning between two multi-qubit quantum states, which lays the ground for our search for a distance measure of clustering topological orders.

\begin{definition} 
\label{def:Def_QPP}
\emph{[Quantum path planning (QPP).]}\ The QPP problem is defined as: 
Given two arbitrary pure $n$-qubit quantum states $\rho_0$ and $\rho_1$, find continuous paths $\rho(s) = U(s) \rho_0 U^{\dagger}(s)$ $(s \in [0, 1])$ that connect $\rho_0$ and $\rho_1$, where the generator $G(s)$ of $U(s)$ is a hermitian operator, \mdf{with the operator norm} $\| \partial_s U(s) \|_{\infty} = \| G(s) \|_{\infty} < + \infty$ $(\forall s \in [0, 1])$. In general, we have that $U(s) \in \mathrm{SU}\left(2^n\right)$ and $i G(s) \in \mathfrak{su}\left(2^n\right)$, with the unitary given by 
\begin{eqnarray}
U(s) &=& {\cal{P}} \ \mathrm{exp} \left( -i \int_0^s G(\tau) \mathrm{d} \tau \right), 
\end{eqnarray}
where $U(s=0) = \mathrm{Id}$ is the identity and ${\cal{P}}$ denotes the path ordering operator. \mdf{In practical implementation, the QPP can involve finding the optimal path under constraints. }
\end{definition}

Note that, for two topologically equivalent gapped quantum states, the generator $G(s)$ in the above QPP must be additionally a geometrically local operator, which underlies why the two quantum states can be transformed into each other via a constant-depth quantum circuit~\cite{ZengArxiv2015} (See the following definition).

\begin{definition}
\label{def:topo_UniPath}
\emph{(Topological equivalence~\cite{ZengPRB2015,ZengArxiv2015}.)}\ 
We say that two gapped ground states $\rho_0$ and $\rho_1$ of local Hamiltonians $H_0$ and $H_1$, respectively, are in the same topological phase if there is a local unitary that maps the two states into each other without closing the energy gap. In other words, there is a smooth path of local Hamiltonians $H(s)$ that connects the two parent Hamiltonians with $H(s = 0) = H_0$ and $H(s = 1) = H_1$ for $s \in [0, 1]$. The Hamiltonian $H(s)$ is gapped above its ground state $\rho(s)$ for all $s \in [0, 1]$, where $\rho(s)$ smoothly connects $\rho_0$ and $\rho_1$, which establishes a smooth path between the two quantum many-body states. In~\cite{BachmannCMP2011}, it is show that such a path can be realized with a smoothly parametrized unitary $U(s)$ whose generator $G(s)$ is also a local operator, $\rho(s) = U(s) \rho_0 U^{\dagger}(s)$, and naturally $\rho_1 = U(s = 1) \rho_0 U^{\dagger}(s = 1)$. This is further elaborated in Sec. II of the Supplementary Information. 
\end{definition}

The above Definition~\ref{def:topo_UniPath} is equivalent to the statement that the two topologically equivalent states can be deformed into each other by a constant-depth quantum circuit~\cite{ZengArxiv2015}, each layer of which can be divided into geometrically local quantum gate operations. Therefore, the two quantum states differ from each other only in entanglement patterns at short-range scales, while they are equivalent in their long-range entanglement structure.

Intuitively, the depth of the quantum circuit quantifies the scale of the change of the quantum entanglement that is caused by applying the circuit on the reference quantum state $\rho_0$. Then, the quantum circuit complexity (QCC) for realizing $U(s = 1)$ can serve as a measure of the algorithmic information for generating the target state $\rho_1$ from a reference state $\rho_0$. As is explained above, this is similar to, and inspired by, the (conditional) Kolmogorov complexity $K(x|y)$, which is the length of the shortest program on a universal computer that outputs a string or dataset $x$ given the access of another string or dataset $y$, and as such represents an algorithmic information distance~\cite{AlgInfoDist} or the shared structure between the two objects~\cite{IlyaNote}. We adopt and adapt this here, using a geometric lower bound of the quantum circuit cost, the Nielsen's QCC $\mathcal{C}_{\mathcal{N}} \left( \rho_0 \rightarrow \rho_1\right)$, as a distance measure for clustering topological states with distinct entanglement patterns, whose definition is given in the following.

\begin{definition} 
\label{def:Nielsen_QCC}
\emph{(Nielsen's QCC.)}\ 
The quantum circuit complexity of the unitary $U(s=1)$ can be evaluated through Nielsen's geometric approach~\cite{NielsenArxiv2005,NielsenScience2006}. For qubit systems, if we write $G(s) = \sum_{\sigma} h_{\sigma} (s) \ \sigma$, where $\sigma \in P_n = \{I, X, Y, Z\}^{\otimes n}$ is a $n$-qubit Pauli operator ($I$ is the $2$-dimensional identity operator and $X, Y, Z$ are the three single-qubit Pauli operators, respectively), and $h_{\sigma} (s)$ denote real coefficients, the Nielsen's quantum circuit complexity of the first order reads 
\begin{eqnarray}
\mathcal{C}_{\mathcal{N}} \left( \rho_0 \rightarrow \rho_1\right) = \underset{h(s)}{\mathrm{inf}} \ \int^1_0 \sum_{\sigma} \left| h_{\sigma} (s) \right| \ \mathrm{d}s. 
\end{eqnarray}
The original Nielsen's QCC can be generalized to systems other than qubits~\cite{HacklJHEP2018,KhanPRD2018,EisertPRL2021}, where an orthonormal operator basis $\{ O_i\}$ replaces the $n$-qubit Pauli operators $P_n$, with the operator norm $\| O_i \|_{\infty} \le 1$ ($\forall i$). Then the definition of Nielsen's QCC above remains unchanged.
\end{definition}

With the Nielsen's QCC of geometrically local quantum circuit as a topological and informational distance measure, the proposed theoretically optimal similarity/ kernel function for the unsupervised manifold learning of topological order is given by   
\begin{eqnarray}
\label{eq:QCC_Kernel_general}
\mathcal{K} \left( \rho_0, \rho_1\right) = \mathrm{exp} \left[-\beta \ \mathcal{C}_{\mathcal{N}} \left( \rho_0 \rightarrow \rho_1\right) \right],
\end{eqnarray}
up to a normalization constant, where $\beta > 0$ is a hyperparameter. The unitarity of the quantum circuit indicates that Nielsen's QCC is symmetric under permutations of the two quantum states, i.e., $\mathcal{C}_{\mathcal{N}} \left( \rho_0 \rightarrow \rho_1\right) = \mathcal{C}_{\mathcal{N}} \left( \rho_1 \rightarrow \rho_0\right)$, which leads to a symmetric kernel in (\ref{eq:QCC_Kernel_general}). \mdf{The similarity function in (\ref{eq:QCC_Kernel_general}) can be used to construct conditional probabilities in the diffusion map~\cite{coifman2005geometric,nadler2006diffusion} and the t-SNE~\cite{Maaten2008t-SNE} manifold learning, or can be used for kernel PCA~\cite{Scholkopf1998Nonlinear,Mika1999Kernel} in the kernel feature space. Physically, Nielsen's QCC can be viewed as a global order parameter of topological quantum orders, and also serves as an ideal topological distance in non-kernel manifold learning such as the metric-MDS~\cite{DemainePMLR2021,SklearnMDS}. } 

\subsection{Unsupervised machine learning of topological order derived from quantum circuit complexity}

While it is in general hard to exactly compute $\mathcal{C}_{\mathcal{N}} \left( \rho_0 \rightarrow \rho_1\right) $ for the QPP, we now establish some approximate substitutes, which both capture essential aspects of Nielsen's QCC and at the same time are easier to implement in practice. The two theorems that we present in the following provide rigorous guarantees for this purpose.

First, we define the quantum Fisher complexity (QFC) related to the QPP, denoted by $\mathcal{C}_{\mathcal{F}} \left( \rho_0 \rightarrow \rho_1\right)$, which coincides with the Bures distance between the two quantum states given by a fidelity measure and can be efficiently calculated in practice.

\begin{definition} 
\label{def:QFC}
\emph{(Quantum Fisher complexity (QFC).)}\
Defining the Uhlmann-Jozsa fidelity between two arbitrary quantum states (pure or mixed) as~\cite{NielsenChuang} 
\begin{eqnarray}
F \left( \rho, \tilde{\rho}\right) = \mathrm{tr} \sqrt{\sqrt{\rho} \tilde{\rho} \sqrt{\rho}}, 
\end{eqnarray}
where $\mathrm{tr}$ denotes the trace operation, then the corresponding squared Bures distance reads~\cite{BraunsteinPRL1994,LiuJPAReview2020,TaddeiPRL2013}
\begin{eqnarray}
D^2_B \left( \rho, \tilde{\rho}\right) = 2 \left[1 - F \left( \rho, \tilde{\rho}\right) \right]. 
\end{eqnarray}

For a family of smoothly parametrized quantum states $\rho(s)$ by $s \in [0, \, 1]$, the quantum Fisher information (QFI) of the quantum state with respect to the parameter $s$ is given by~\cite{Helstrom,Holevo,MaPhysRep2011,TaddeiPRL2013,LiuJPAReview2020} $\mathcal{F}_Q \left( s \right) = \mathrm{tr} \left[\rho \left( s \right) \mathcal{L}^2\right]$, with the symmetric logarithmic derivative (SLD) $\mathcal{L}$ determined by $\partial_s \rho \left( s \right) = \frac{1}{2} \left[\rho \left( s \right) \mathcal{L} + \mathcal{L} \rho \left( s \right) \right]$. The QFI, which plays a central role in quantum parameter estimation and which is also used for characterizing multipartite entanglement and topological states~\cite{YuRanPRL2017}, can be related to the infinitestmal Bures distance~\cite{BraunsteinPRL1994} with  
\begin{eqnarray}
\label{eq:Bures_QFI}
\frac{1}{4} {\cal{F}}_Q (s) \mathrm{d}^2s = D^2_B \left[ \rho(s), \rho(s+\mathrm{d}s)\right],  
\end{eqnarray}
representing a susceptibility of the fidelity.

\medskip
The QFC of the target quantum state $\rho_1$ with respect to the reference state $\rho_0$ is given by~\cite{ChapmanPRL2018}
\begin{eqnarray}
\mathcal{C}_{\mathcal{F}} \left( \rho_0 \rightarrow \rho_1\right) = \underset{h(s)}{\mathrm{inf}} \ \frac{1}{2} \int^1_0 \sqrt{ {\mathcal{F}}_Q (s) }\ \ \mathrm{d}s, 
\end{eqnarray}
which represents the optimal Bures distance between the two quantum states~\cite{TaddeiPRL2013}, reflecting the overall fidelity variation, as indicated by (\ref{eq:Bures_QFI}). Note that we have a symmetric QFC under permutations of its inputs, i.e., $\mathcal{C}_{\mathcal{F}} \left( \rho_0 \rightarrow \rho_1\right) = \mathcal{C}_{\mathcal{F}} \left( \rho_1 \rightarrow \rho_0\right)$.
\end{definition}

\bigskip
With the QFC and the Bures distance, we are ready to formulate the following theorem. 

\begin{theorem}
\label{thm:QCC_Fidelity}
\emph{(Quantum circuit complexity upper bounds \mdf{quantum Fisher complexity and Bures distance}.)} \ 
The QCC for the QPP from $\rho_0$ to $\rho_1$ is lower bounded by the QFC and the Bures distance, 
\begin{eqnarray}
\label{eq:FisherNielsenComplexity}
\mathcal{C}_{\mathcal{N}} \left( \rho_0 \rightarrow \rho_1\right) &\ge& {\cal{C}}_{\mathcal{F}} \left( \rho_0 \rightarrow \rho_1\right)  \nonumber  \\
&\ge& \frac{D_B \left( \rho_0, \rho_1\right)}{\sqrt{2}}.
\end{eqnarray}
Furthermore, the QCC of a geometrically local quantum circuit for the same QPP is approximately lower bounded by 
\begin{eqnarray}
\label{eq:QCC_Fidelity}
\mathcal{C}_{\mathcal{N}} \left( \rho_0 \rightarrow \rho_1\right) \gtrsim \frac{1}{\sqrt{2}} \, \underset{\Delta}{\sum} \, D_B \left[ \rho_0 \left( \Delta \right), \rho_1 \left( \Delta \right) \right],
\end{eqnarray}
where $\rho \left( \Delta \right)$ is the reduced density matrix supported on the subsystem $\Delta$ of constant size, and the summation goes over non-overlapping subsystems which, together with their neighboring environments, cover the whole system.
\end{theorem}

The proof of Theorem~\refeq{thm:QCC_Fidelity} is given in Sec. I of the Supplementary Information, \mdf{and an asymptotic and tightness analysis of the bounds in Theorem~\refeq{thm:QCC_Fidelity} is provided in Sec. III.} Note that, without the constraint of geometric locality of the quantum circuit, the overall Bures distance in (\ref{eq:FisherNielsenComplexity}) is equivalent to the $L_2$-norm distance between the two density matrices, which has been shown not to be a suitable distance measure for topological phases~\cite{ChePRB2020}. 

Under the geometric locality of the quantum circuit, we exploit Theorem~\ref{thm:QCC_Fidelity} to construct a fidelity-based kernel. The Uhlmann-Jozsa fidelity between the $r$-body reduced density matrices with $r = \{1, 2, \cdots, R \}$ can be used for this purpose, where $R$ is a scale of cutoff. For instance, for qubits hosted on a lattice $\Lambda = [1, \ L]^D$ of lattice size $L$ and dimension $D$, the proposed kernel based on fidelity reads (up to some normalization constant)
\begin{eqnarray}
\label{eq:K_F}
\mathcal{K}_{\mathrm{F}} \left( \rho, \tilde{\rho}\right) = \mathrm{exp} \left\{ \beta \ \underset{r = 1}{\sum^R} \omega_r \underset{\Delta \in P_r (\Lambda)}{\sum} F \left[\rho (\Delta), \ \tilde{\rho} (\Delta)\right] \right\},
\end{eqnarray}
where $P_r (\Lambda) = \{\Delta | \Delta \subset \Lambda \ \text{and} \ \left| \Delta \right| = r \}$, with $\left| \Delta \right|$ being the size (cardinality) of the sublattice $\Delta$; $\rho (\Delta)  = \mathrm{tr}_{(\Lambda - \Delta)} \rho$ is the reduced density matrix supported on $\Delta$; $\omega_r$ is a weight function; and $F(\cdot, \cdot)$ is the Uhlmann-Jozsa fidelity. 
While the dimension of the reduced density matrix depends exponentially on its size $r$, this fidelity-based kernel can be efficiently estimated in practical implementations for $R = \mathrm{const.}$ (For instance, a small $r = 2$ is already sufficient in many relevant applications. See Theorem~\ref{thm:QCC_Fidelity} and the section on numerical experiments below.)

\bigskip
The second approximation of the Nielsen's QCC that we propose focuses on the entanglement change. Denoting by $\rho$ a pure density matrix of a $n$-qubit chain (rearranged from an arbitrary spatial geometry following a specified order), and $A_k = \{ 1, 2, ..., k \} \subset [n]$ a subset of the system, which is obtained from a cut at $k \in [1, \ n-1]$, then the reduced density matrix supported on $A_k$ reads $\rho (A_k) = \mathrm{tr}_{B_k} \rho$, where $B_k = \{ k+1, k+2, ..., n \}$. With the entanglement entropy over the cut $k$ given by 
\begin{equation}
S_k (\rho) = \mathrm{tr} \left[ \rho (A_k) \ln \rho (A_k) \right],
\end{equation}
we have the following theorem: 
\begin{theorem}
\label{thm:QCC_Entanglement}
\emph{[Quantum circuit complexity upper bounds \mdf{entanglement-profile distance} (A generalized version of Observation 1 in~\cite{EisertPRL2021}).]} 
The QCC for the QPP from $\rho_0$ to $\rho_1$ with a geometrically local quantum circuit (with respect to the $n$-qubit chain) is lower bounded by 
\begin{equation}
\label{eq:QCC_Ent}
\mathcal{C}_{\mathcal{N}} \left( \rho_0 \rightarrow \rho_1\right) \ge \frac{c}{\left(n - 1 \right)} \sum^{n-1}_{k=1} \left| S_k (\rho_1) - S_k (\rho_0) \right|,
\end{equation}
for some constant $c > 0$.
\end{theorem}
The proof of Theorem~\refeq{thm:QCC_Entanglement} is given in Sec. I of the Supplementary Information, \mdf{and an asymptotic and tightness analysis of the bounds in Theorem~\refeq{thm:QCC_Entanglement} is provided in Sec. III.} Theorem~\ref{thm:QCC_Entanglement} can be readily generalized to situations where the qubits are arranged in higher spatial dimensions and other geometries, provided that the quantum circuit is applied in a geometrically local manner accordingly. Nielsen's QCC is then lower bounded by the entanglement change distributed over suitable cuts covering non-trivial bonds of the system.   

Theorem~\ref{thm:QCC_Entanglement} inspires an entanglement-based kernel, given by
\begin{eqnarray}
\label{eq:K_E}
\mathcal{K}_{\mathrm{E}} \left( \rho, \tilde{\rho}\right) = \mathrm{exp} \left( -\frac{\beta}{n} \sum^{n-1}_{k=1} \left| S_k (\rho) - S_k (\tilde{\rho}) \right| \right),
\end{eqnarray}
which is particularly useful when the entanglement profile over nontrivial bonds can be efficiently estimated. This is, for instance, the case in tensor-network simulations of quantum many-body states such as the density matrix renormalization group (DMRG)~\cite{WhitePRL1992,SchollwAnnPhys2011}, or in experimentally prepared quantum states measured via classical shadow tomography~\cite{HuangNatPhys2020}. 

Both lower bounds of Nielsen's QCC in (\ref{eq:QCC_Fidelity}) and (\ref{eq:QCC_Ent}) potentially capture important information about the topological order. The fidelity-based kernel with reduced density matrices of constant size can be efficiently calculated and benchmarked (For instance, with classical shadow representations, see the following section). The entanglement-based kernel is more stringent and contains more information about quantum entanglement at multiple scales.

\subsection{Relation to classical shadows}
\label{sec:classical_shadow}
Classical shadow (CS) tomography is an experimentally friendly approach to access many properties of quantum many-body states with few-shot measurements~\cite{HuangNatPhys2020}. In this formalism, the quantum state is approximated by its CS estimator, 
\begin{eqnarray}
\rho \approx S_T\left(\rho\right) = \frac{1}{T} \sum_{t=1}^T \sigma^{(t)}_1 \otimes \sigma^{(t)}_2 \cdots \otimes \sigma^{(t)}_n,
\end{eqnarray}
in a total of $T$ measurements, where $\sigma^{(t)}_i$ is the local CS representation under the Pauli measurement for the $i$-th qubit in the $t$-th measurement. The shadow kernel suggested in~\cite{HuangScience2022Manybody}, given by
\begin{eqnarray}
\label{eq:K_CS}
\mathcal{K}_{\mathrm{CS}} \left[ S_T\left(\rho\right), S_T\left(\tilde{\rho} \right) \right] = \mathrm{exp} \left\{ \frac{\beta}{T^2} \underset{t, \ t^{\prime} = 1}{\sum^T} \ \mathrm{exp} \left[ \frac{\nu}{n} \underset{i = 1}{\sum^n} \mathrm{tr}\left(\sigma^{(t)}_i \tilde{\sigma}^{(t)}_i \right) \right] \right \},
\end{eqnarray}
can be understood as a special case of the fidelity-based kernel in (\ref{eq:K_F}), if the quantum state $\rho$ and the $r$-body reduced density matrices are represented by their classical shadows, if the weight function is taken to be $\omega_r = \nu^r / r!$, and if we take the limit $R \rightarrow \infty$, where $\beta > 0$ and $\nu > 0$ are constant hyperparameters. In this case, density matrices with small values of $r$ dominate in the summation, and randomized measurements reduce the complexity in calculating fidelity-based kernels.
\begin{figure*}[t]
\centerline{\includegraphics[height=4.6in,width=6in,clip]{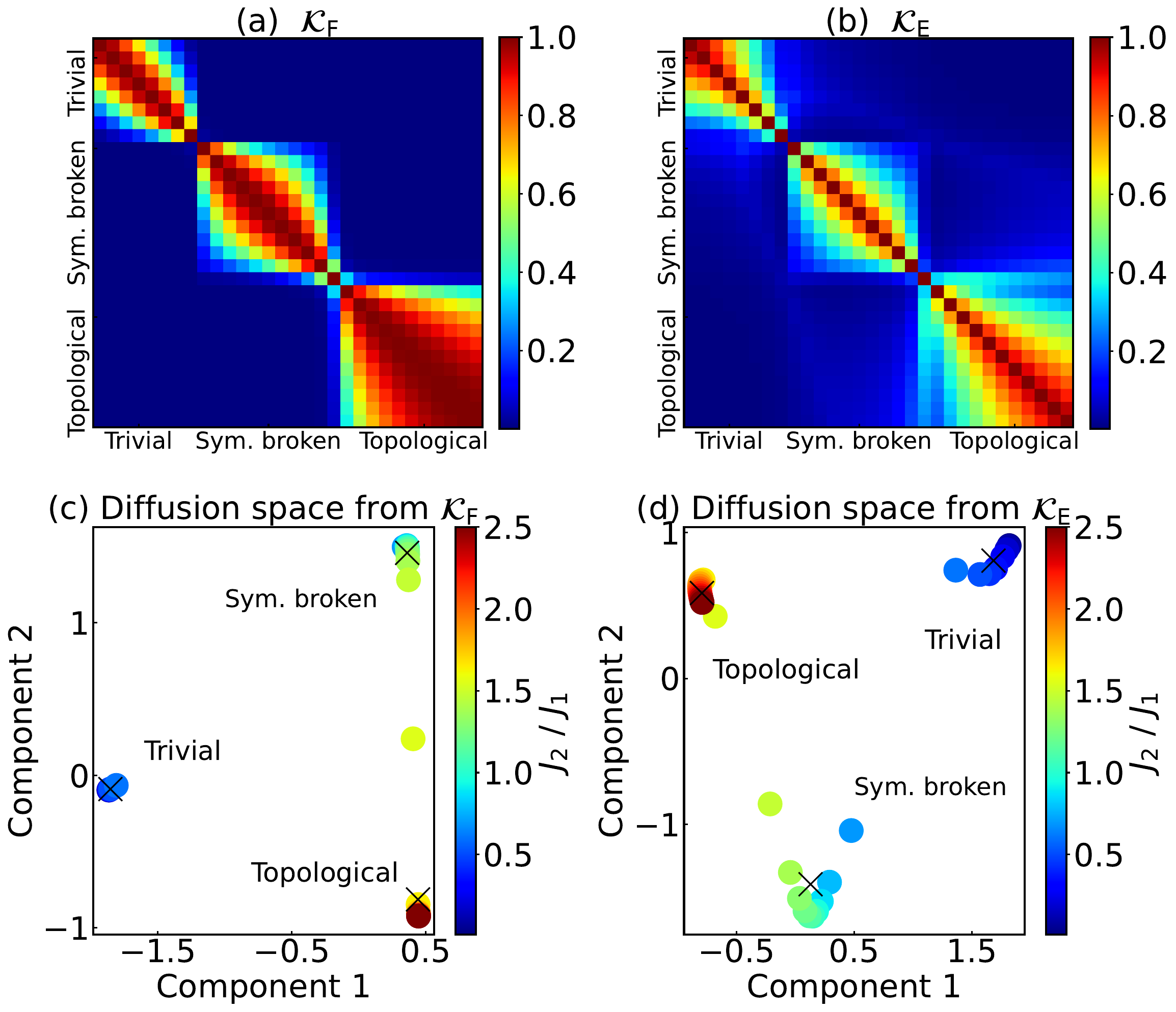}}
\caption{\textbf{Unsupervised manifold learning of the bond-alternating XXZ qubit chain with fidelity- and entanglement-based informational distances.} (a) and (b) show the fidelity- and entanglement-based kernels, $\mathcal{K}_{\mathrm{F}} \left( \rho, \tilde{\rho}\right)$ and $\mathcal{K}_{\mathrm{E}} \left( \rho, \tilde{\rho}\right)$, respectively. One sees clearly three clusters in the heat map, corresponding to three distinct quantum phases of the model (trivial, symmetry broken, and topological) in a range of the model parameter $J_2/J_1 \in (0,\ 2.5]$, which is represented by the two axes of (a) and (b), respectively. The colorbar encodes the value of the normalized kernel. (c) and (d) are the corresponding two-dimensional ($\mathrm{2D}$) nonlinear representations of the samples, through the diffusion map algorithm [plotted are the second and the third dimensions in the diffusion space (Components 1 and 2, respectively), after normalization by the standard deviation]. The colorbar encodes the value of $J_2/J_1$, and the cross in each plot indicates the centre of the cluster found by a $k$-means clustering algorithm. \mdf{The outlier in (c) with the value of $J_1 / J_2 \approx 1.5$ is the critical phase due to the finite-size effect and the resulted  finite-width of the critical region. We use $n = 151$} qubits, and the DMRG [with the singular value decomposition (SVD) cutoff $10^{-10}$ and the maximal energy error $10^{-10}$] to determine the ground state. We set $h_0 = 0$ and $\delta = 3$, while $N = 30$ samples are drawn uniformly in the parameter range of $J_2/J_1 \in (0,\ 2.5]$ in an ordered manner. The hyperparameter of the kernel is $\beta = 50.0 $ for (a) and (c); and \mdf{$\beta = 10.0$} for (b) and (d). The entanglement profile is accessible in the DMRG calculation with the matrix product state representation. In the fidelity-based kernel, \mdf{we use $n$ geometrically local (nearest-neighbour) two-body reduced density matrices to cover the system,} i.e., we only use terms with $r = 2$ and $\omega_{r=2} = 1/n$ in (\ref{eq:K_F}).
}
\label{fig:XXZ_Model}
\end{figure*}
\subsection{Numerical experiments}
\label{sec:numeric_experiments}
In this section, we provide numerical demonstrations for the developed theory of unsupervised machine learning of topological order based on the QCC. 

\subsubsection{The bond-alternating XXZ spin chain} 

We begin with a benchmark model, the bond-alternating XXZ spin-$\frac{1}{2}$ chain~\cite{PollmannPRB2012,ElbenSA2020,HuangScience2022Manybody}, whose Hamiltonian is given by 
\begin{eqnarray}
\label{eq:XXZ_Hamiltonian}
H_{\mathrm{XXZ}} = \underset{i = 1}{\sum^{n-1}} J_i \left(X_i X_{i+1} + Y_i Y_{i+1} + \delta Z_i Z_{i+1} \right) - h_0 \underset{i = 1}{\sum^{n}} Z_i,
\end{eqnarray}
where $J_i = J_1$ if $i$ is odd and $J_i = J_2$ if $i$ is even, respectively; $\delta$ is a detuning parameter and $h_0$ denotes the strength of the external field.

In Fig.~\ref{fig:XXZ_Model}, we present a comparison between the fidelity- and entanglement-based kernels for this model, respectively, together with the two-dimensional representation in the diffusion space from manifold learning of the diffusion map algorithm~\cite{coifman2005geometric,nadler2006diffusion,Rodriguez-NievaNatPhys2019,ChePRB2020,ScheurerPRL2020}. The ground-state properties (fidelities and entanglements) are obtained from DMRG calculations of the model. 
Fixing $h_0 = 0$, $\delta = 3$ and varying the value of $J_2/J_1$, we find that the unsupervised learning produces distinct clusters that can be associated with three phases (trivial, symmetry broken, and topological), in good agreement with independent calculations of topological invariants~\cite{PollmannPRB2012,ElbenSA2020,HuangScience2022Manybody}. More details can be found in the caption of Fig.~\ref{fig:XXZ_Model}. While we are ultimately aiming at the unsupervised learning of topological order in the presence of non-local quantum entanglement, Fig.~\ref{fig:XXZ_Model} shows that the fidelity- and entanglement-based kernels also perform well when clustering symmetry-broken phases and symmetry-protected topological orders~\cite{ElbenSA2020} that are characterized by short-range entanglement. 

\begin{figure*}[t]
\centerline{\includegraphics[height=4.8in,width=6in,clip]{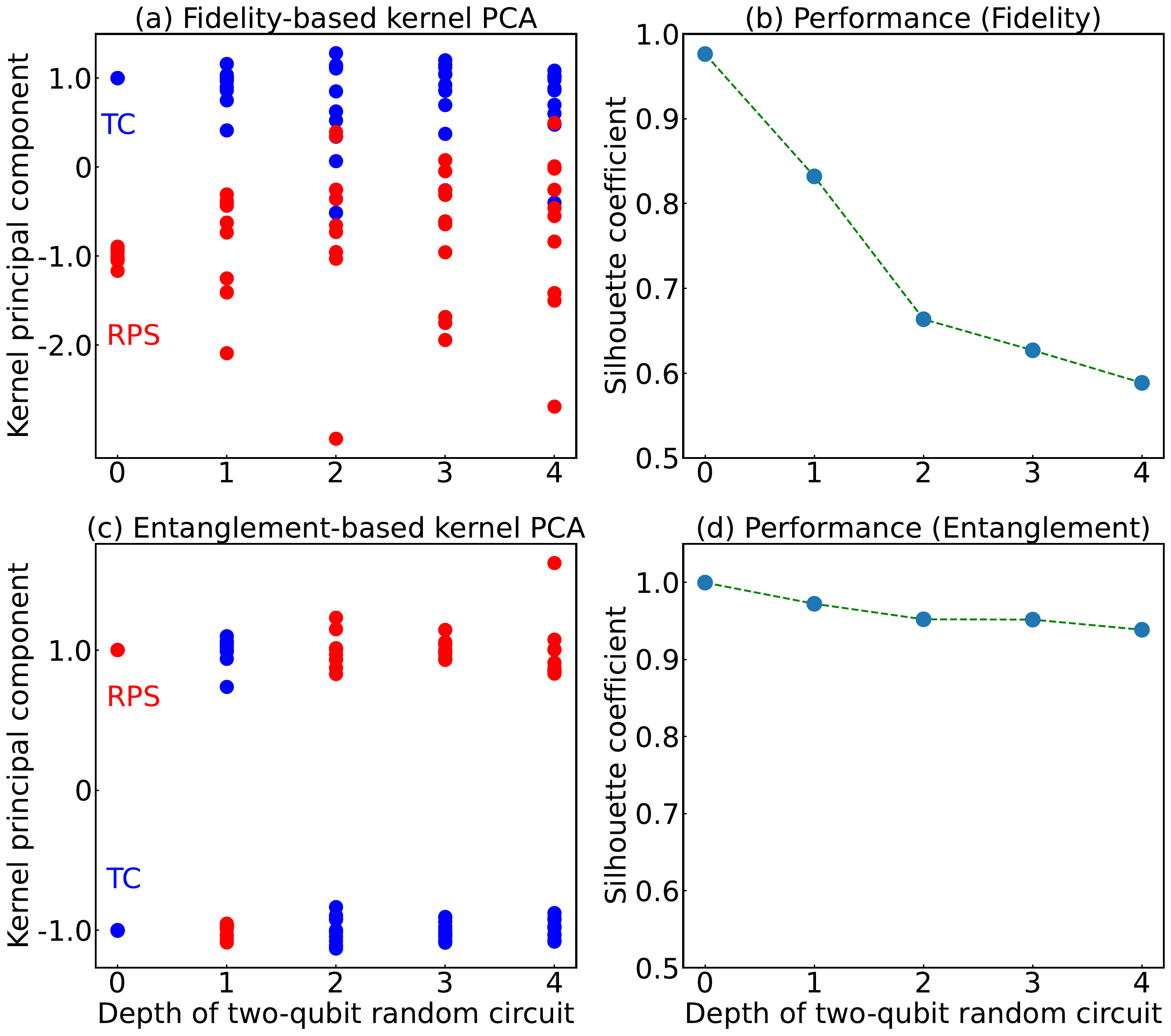}}
\caption{\textbf{Unsupervised clustering of the ground state of Kitaev's toric code (TC) [blue dots in sub-figures (a) and (c)] and random product states (RPS) [red dots in sub-figures (a) and (c)] without or with applying two-qubit random unitaries, via fidelity- and entanglement-based kernels, respectively.} (a) and (c) show a one-dimensional ($\mathrm{1D}$) representation of the data set in terms of the kernel's principal components (i.e., via the algorithm of kernel PCA with shifting the center to zero and normalizing by the standard deviation), by applying repeatedly different depths of random two-qubit unitaries with respect to the Haar measure~\cite{tenpy} to both the toric code and the RPS. (a) and (c) utilize the fidelity- and entanglement-based kernels, $\mathcal{K}_{\mathrm{F}} \left( \rho, \tilde{\rho}\right)$ and $\mathcal{K}_{\mathrm{E}} \left( \rho, \tilde{\rho}\right)$, respectively. \mdf{(b) and (d) plot the degradation of clustering performance under Haar random noises for results in (a) and (c), respectively, with the silhouette coefficient~\cite{Sklearn_Silhouette} (which takes the value one for best clustering) as a benchmark metric. The four subplots share the same horizontal axis.} We use $N = 20$ samples in total, with 10 samples of the toric code (blue dots) and 10 samples of the RPS (random spin ups and downs of $n$ qubits) (red dots). We use $n = 32$ qubits (on a lattice of size $4 \times 4$ with toric boundary condtions; i.e., with a small code distance of $4$) for this proof-of-principle demonstration, and the DMRG (with the SVD cutoff $10^{-10}$ and the maximal energy error $10^{-10}$) to solve for the ground state of the toric code and, at the same time, to extract the entanglement profile. The hyperparameter of the kernel is $\beta = 0.1$ for (a) and $\beta = 2.0$ for (c). With the fidelity-based kernel, we randomly sample $n$ two-body reduced density matrices~\cite{Note_RDM}, i.e., we only use terms with $r = 2$ and $\omega_{r=2} = 1/n$ in (\ref{eq:K_F}). The result shows that, in the absence of noise (i.e. with circuit depth zero), both kernels types perform well in clustering the toric code and the RPS, while the entanglement-based kernel in (c) is found to be more robust to Haar random two-qubit unitaries [see (b) and (d)]. 
}
\label{fig:TC_Model}
\end{figure*}

\subsubsection{Kitaev's toric code}  
As a second demonstration, we cluster the ground state of Kitaev's toric code with random product states, where the former has topological quantum order while the latter is topologically trivial. The toric code Hamiltonian defined on a $\mathrm{2D}$ lattice of size $L_x \times L_y$ is given by~\cite{KitaevAnnPhys2003} 
\begin{eqnarray}
\label{eq:TC_Hamiltonian}
H_{\mathrm{TC}} = - \underset{v}{\sum} A_v - \underset{p}{\sum} B_p,
\end{eqnarray}
where $v$ denotes a vertex and $p$ is the face of the lattice, respectively; $A_v = \underset{j \in \mathrm{star}(v)}{\prod} X_j$ and $B_p = \underset{j \in \mathrm{boundary}(p)}{\prod} Z_j$ are the Kitaev's star and face operators, respectively. The number of qubits for this model is $n = 2 L_x L_y$. The toric code ground state is gapped and exhibits area-law entanglement~\cite{KitaevPRL2006}. Therefore, for the purpose of a proof-of-principle demonstration and in order to obtain the entanglement profile for the kernel with a small system size, we use a MPS representation and DMRG to find a reasonably accurate solution for the ground state. For instance, numerical experiments show that, for a $4 \times 4$ lattice with periodic boundary conditions in both dimensions, a maximal MPS bond dimension of $64$ is sufficient to produce accurate results. 

In Fig.~\ref{fig:TC_Model} we plot a comparison between the unsupervised machine learning of the toric code ground state (blue dots) and the random product states (RPS) (red dots) of qubits (i.e., random bit strings of spin ups and downs), with fidelity- and entanglement-based kernels, respectively. Both kernels perform well in clustering the topologically ordered quantum state and the trivial phase, while the entanglement-based one is more robust under the application of two-qubit Haar random unitaries to both the toric code and the RPS. \mdf{This follows from the comparison between Fig.~\ref{fig:TC_Model} (b) and (d), where the silhouette coefficient~\cite{Sklearn_Silhouette}, which takes the value one for best clustering, is used for illustrating the performance degradation of the clustering results in Fig.~\ref{fig:TC_Model} (a) and (c), respectively, under Haar noises. The entanglement created by applying the two-qubit random circuit grows exponentially with the circuit depth. More details about the plot can be found in the caption of Fig.~\ref{fig:TC_Model}. For more discussions on the effects of hardware noises and applied random quantum circuits, we refer the reader to Sec. V of the Supplementary Information. In addition, in Sec. VI of the Supplementary Information (see Fig. S1), we also use metric-MDS (which does not require a kernel formulation) based on the Bures and entanglement-profile distance metrics in (\ref{eq:QCC_Fidelity}) and (\ref{eq:QCC_Ent}), respectively, for unsupervised manifold clustering of the toric code and the product state. Similar conclusions can be obtained from Fig. S1. }

\begin{figure*}[b]
\centerline{\includegraphics[height=4.5in,width=4.5in,clip]{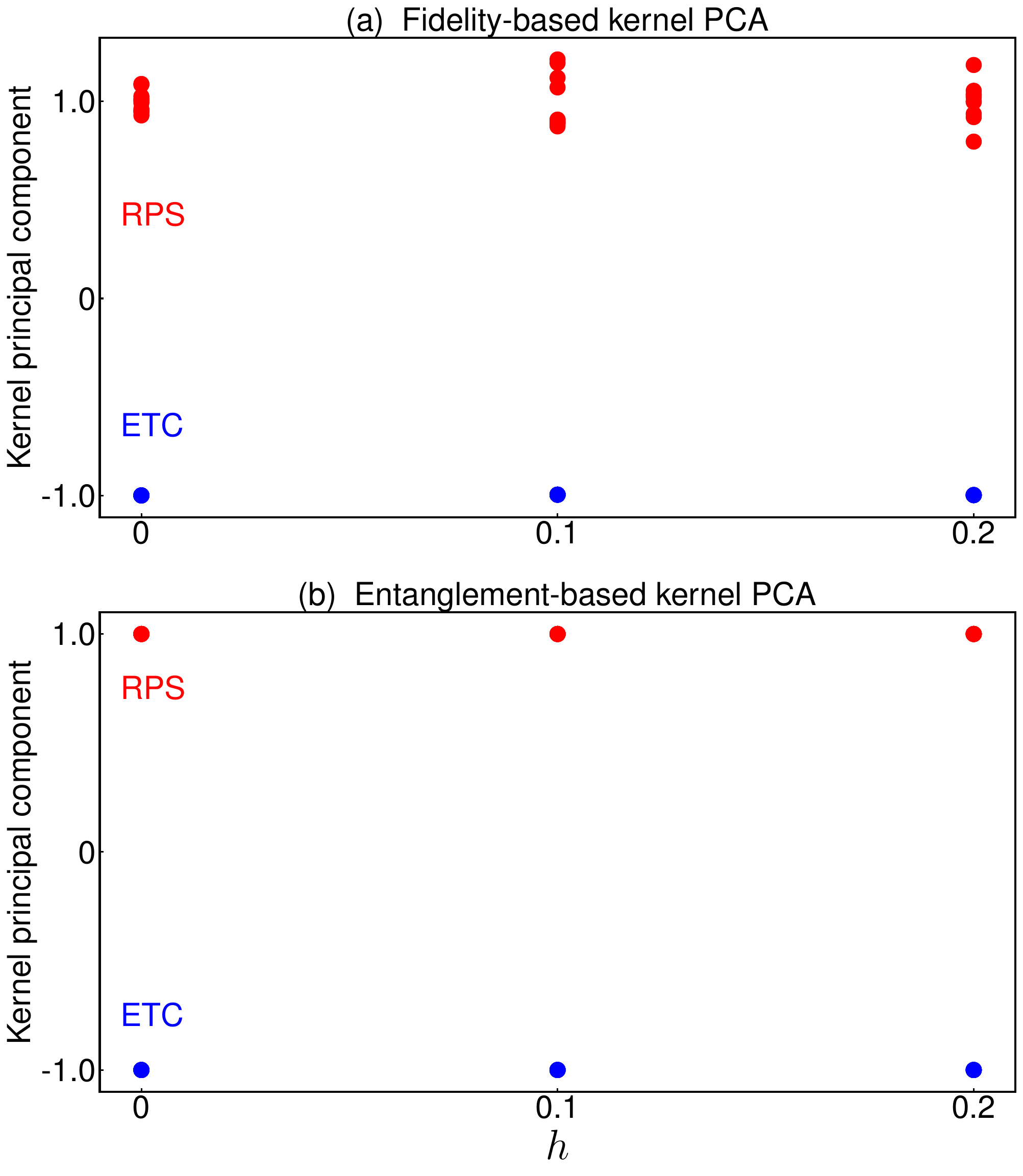}}
\caption{\textbf{Unsupervised clustering of the ground state of the extended toric code (ETC) (blue dots) and random product states (RPS) (red dots) without or with applying local perturbations, via fidelity- and entanglement-based kernels, respectively.} (a) and (b) show a one-dimensional ($\mathrm{1D}$) representation of the data set in terms of the kernel's principal components (i.e., via the algorithm of kernel PCA with shifting the center to zero and normalizing by the standard deviation), by applying different field strengths of local perturbations ($h$) to the ETC. (a) and (b) utilize the fidelity- and entanglement-based kernels, $\mathcal{K}_{\mathrm{F}} \left( \rho, \tilde{\rho}\right)$ and $\mathcal{K}_{\mathrm{E}} \left( \rho, \tilde{\rho}\right)$, respectively. The two subplots share the same horizontal axis. We use $N = 20$ samples in total, with 10 samples of the ETC (blue dots) and 10 samples of the RPS (random spin ups and downs of $n$ qubits) (red dots). The boundary conditions are given by an infinite cylinder geometry, where it is periodic in the $y$ direction with a circumference of $L_y = 4$; In the $x$-direction it is infinite, and we choose a unit cell of width $L_x = 2$. The MPS-based DMRG (with the SVD cutoff $10^{-10}$ and the maximal energy error $10^{-10}$) is used to solve for the ground state of the ETC and, at the same time, to extract the entanglement profile for both zero and nonzero values of $h$. The hyperparameter of the kernel is $\beta = 0.2$ for (a) and $\beta = 2.0$ for (b). With the fidelity-based kernel, we randomly sample $n$ two-body reduced density matrices, i.e., we only use terms with $r = 2$ and $\omega_{r=2} = 1/n$ in (\ref{eq:K_F}). The result shows that the topological clustering is robust to local perturbations.
}
\label{fig:TC_Ext}
\end{figure*}

\subsubsection{The extended toric code}  
The original toric code is defined on a torus with periodic boundary conditions in both spatial dimensions. In contrast, the extended toric code (ETC) is defined on an infinite cylinder (e.g., the cylinder axis is along the $x$ direction and the $y$ direction is periodic with circumference $L_y$), and additionally features the Wilson and t'Hooft loop operators $W$ and $H$, respectively. The corresponding Hamiltonian is given by~\cite{tenpy} 
\begin{eqnarray}
\label{eq:ETC_Hamiltonian}
H_{\mathrm{ETC}} = H_{\mathrm{TC}} - J_{\mathrm{W}} W - J_{\mathrm{H}} H - h \underset{i = 1}{\sum^{n}} Z_i,
\end{eqnarray}
where $H_{\mathrm{TC}}$ is the original toric code Hamiltonian in (\ref{eq:TC_Hamiltonian}), and $h$ is a (uniform) local field strength. The loop operators are defined as products 
\begin{eqnarray}
W = \underset{i \in E_{v}}{\prod} Z_i, \, \, \, \, \,  H = \underset{i \in E_{h}}{\prod} X_i
\end{eqnarray}
around the cylinder through vertical edges ($E_v$) and horizontal edges ($E_h$) of the lattice graph, respectively. 

In Fig.~\ref{fig:TC_Ext}, by fixing the values of $J_W = +1$ and $J_H = -1$, which stabilizes the ground state into one of the four degenerate sectors, and by varying the strength of the local perturbation field $h$, we plot a comparison between the unsupervised machine learning of the ETC ground state (blue dots) and the random product states (RPS) (red dots) of qubits (i.e., random bit strings of spin ups and downs), with fidelity- and entanglement-based kernels, respectively. Note that, for finite values of $h$, the ground states are no longer exactly solvable. The required maximal MPS bond dimension increases rapidly with larger values of $h$ even for small system sizes (For the system size and the parameter range we adopted in Fig.~\ref{fig:TC_Ext}, a maximal bond dimension of $600$ is sufficient to produce accurate results). The topological clustering shows its stability against local perturbations. More details about the plot can be found in the caption of Fig.~\ref{fig:TC_Ext}.

The DMRG ground-state search in the numerical experiments was performed using the TeNPy Library~\cite{tenpy}, and some example codes there are reused for preparing the dataset of our machine learning model. 

\section{Discussion}
\label{sec:outlook}
In summary, we used quantum circuit complexity, a vital concept in quantum computation, to understand and to build viable distance measures and kernels for the unsupervised machine learning of topological order, targeting interpretability and generalizability. By resorting to Nielsen's quantum circuit complexity, two easier-to-implement \mdf{distance metrics} based on fidelity and entanglement were proposed and their effectiveness numerically verified. \mdf{Equipped with our topologically informational distance metrics, kernel-based and non-kernel manifold learning can be applied for interpretable phase clustering of quantum many-body systems. }

The shadow kernel learning of~\cite{HuangScience2022Manybody} can be naturally explained by our method derived from the Nielsen's QCC, when the fidelity-based distance and classical shadow representation of the quantum state are used. The alternative approach based on the entanglement-entropy profile encodes more information about entanglement patterns in topological orders, and thus exhibits superior robustness against two-qubit random noises; moreover, they are more interpretable in terms of the relationship between topological quantum orders and long-range entanglement. 
Recent theoretical studies~\cite{VermerschPRX2024,VermerschPRXQ2024} show the promise of efficient measurement of many-body entropies, which indicates that the entanglement-based distance can be calculated efficiently from experimental data. This implies that our method can be readily applied to experimentally prepared multi-qubit quantum states; for instance, data obtained from near-term quantum computing hardware. 
When limited to SPT order and compared to the recent results in~\cite{WuNC2024} about neural-network learning topological orders of the bound-alternating XXZ model, our method is fully unsupervised and results in sharper phase boundaries in clustering quantum phases of this model while providing good interpretability. 

In future work, the kernels proposed here may be also used for supervised learning tasks, such as classification or regression of topological phases, with an $\mathcal{O}(1)$ sample complexity. The nonlinear feature vectors (e.g., random Fourier features) given by the entanglement kernel can be used to learn about properties related to entanglement structures of the quantum many-body state. Also, in the demonstrated examples, so far we focused on pure and gapped ground states. Possible extensions could be applications in gappless, topological quantum criticalities, e.g., with the multiscale entanglement renormalization ansatz (MERA)~\cite{VidalPRL2008}, or in the unsupervised machine learning of entanglement transitions~\cite{IppolitiPRX2021,GranetPRL2023}, as well as mixed-state topological order~\cite{FanPRXQ2024,EllisonPRXQ2025,SohalPRXQ2025,WangPRXQ2025,MaPRXQ2025}.

Our current approach is primarily based on the classical machine learning of topological quantum order by exploiting Nielsen's QCC as an informational distance. It might also be interesting to explore quantum machine learning~\cite{Biamonte2017QML,BartkiewiczSciRep2020} in terms of parametrized quantum circuits (e.g., ~\cite{Petrarxiv2024}), which may find approximately the minimal quantum circuit matching the reference and target quantum states in the data set. The combination with reinforcement learning for quantum state generation~\cite{Barr2020,YaoPRX2021,ZengPRL2023} can be also promising. We also note a very recent work~\cite{LakePRXQ2025} which uses exact quantum algorithms for topological quantum phase recognition, while restricted to one-dimensional SPT phases. 

We believe that the results presented here make, both conceptually and technically, an important step toward an interpretable and generalizable theory of unsupervised machine learning of topological order, and as such will boost the interplay between the research fields of (interpretable) machine learning phases of matter, quantum complexity, quantum parameter estimation, and quantum computation.

\section{Methods}
\label{sec:Methods}
The pivot of our method is an informational distance metric for topological dissimilarities, leveraging Nielsen's geometric QCC of mapping two quantum many-body states as a primitive distance metric for manifold learning. This can result in both good interpretability and generalizability for unsupervised machine learning of topological quantum order. The theoretical toolbox includes differential manifolds, quantum Fisher information, entanglement-growth rate, concept of topological equivalence, and quantum circuit computation, etc. 
By bridging conceptual power to practical implementation, Theorems 1 and 2 lead to approximate substitutes of the QCC, i.e., the Bures distance (fidelity change) and the entanglement-entropy distance, respectively, which preserve partial but important information of topological dissimilarity and which are intuitively more comprehensive and more accessible in simulations and measurements. As a comparison, we find that the entanglement-based manifold learning contains more topological information and performs better than the Bures-distance based one under Haar random noises. Regarding specific manifold-learning algorithms, we use the well-established framework with diffusion map~\cite{coifman2005geometric,nadler2006diffusion}, kernel PCA~\cite{Scholkopf1998Nonlinear,Mika1999Kernel}, and 
metric-MDS~\cite{DemainePMLR2021,SklearnMDS} for demonstrations in the current paper. We believe that other manifold-learning algorithms such as the t-SNE~\cite{Maaten2008t-SNE} and the Isomap~\cite{TenenbaumScience2000}, equipped with our topologically informational distances, will give similar results. 

In the Supplementary Information, we provide, respectively, the details for the proofs of the theorems, a scaling and tightness analysis of the proved bounds, a complexity and scaling analysis of the method with respect to the system size (or the number of qubits $n$) and the sample size $N$, effects of noises and random circuits, and a (non-kernel) metric-MDS manifold learning of the toric-code. 

\section{Acknowledgments} 
C.G. is partially supported by RIKEN Incentive Research Projects.
X.W. is supported by the Hangzhou Joint Fund of the Zhejiang Provincial Natural Science Foundation of China under Grant
No. LHZSD24A050001, and the Science Foundation of Zhejiang Sci-Tech University (Grants No. 23062088-Y, and No. 23062181-Y), and NSFC (No. 24062271-A).  
F.N. is supported in part by: the Japan Science and Technology Agency (JST) [via the CREST Quantum Frontiers program Grant No. JPMJCR24I2, the Quantum Leap Flagship Program (Q-LEAP), and the Moonshot R\&D Grant Number JPMJMS2061], and the Office of Naval Research (ONR) Global (via Grant No. N62909-23-1-2074).



\clearpage{}

\setcounter{equation}{0} \setcounter{figure}{0}
\setcounter{section}{0} 
\renewcommand{\theequation}{S\arabic{equation}}
\renewcommand{\thefigure}{S\arabic{figure}}

\makeatletter
\def\@hangfrom@section#1#2#3{\@hangfrom{#1#2#3}}
\makeatother

\begin{center}{\Large SUPPLEMENTARY INFORMATION for \\~ \\ ``Quantum circuit complexity and unsupervised machine learning of topological order"}

\vskip 2ex
{Yanming Che,$^{1, \ 2}$ \  Clemens Gneiting,$^{2}$ \ Xiaoguang Wang,$^{3}$ \ and Franco Nori$^{1, \ 2}$}\\
\vskip 1ex
\small \textit{$^1$Department of Physics, University of Michigan, Ann Arbor, Michigan 48109-1040, USA \\
$^2$Center for Quantum Computing, RIKEN, Wako-shi, Saitama 351-0198, Japan \\ 
$^3$Department of Physics, Zhejiang Sci-Tech University, Hangzhou 310018, China
}
\end{center}

\date{\today}


\bigskip \bigskip \bigskip

\noindent 
\begin{center}\textbf{ \Large{} Table of contents }{\Large\par}
\end{center}

\bigskip

\noindent \textbf{\ref{sec:ProofThms}.~~\hyperref[sec:ProofThms]{Proof of Theorems}} \dotfill\textbf{\pageref{sec:ProofThms}}
\medskip

\noindent \textbf{\ref{sec:AQC}.~~\hyperref[sec:AQC]{Topological equivalence of gapped ground states and quantum circuit complexity of adiabatic quantum computation}} \dotfill\textbf{\pageref{sec:AQC}}
\medskip

\noindent \textbf{\ref{sec:Tightness_analysis}.~~\hyperref[sec:Tightness_analysis]{Asymptotic and tightness analysis of the bounds in Theorems 1 and 2}} \dotfill\textbf{\pageref{sec:Tightness_analysis}}
\medskip

\noindent \textbf{\ref{sec:Scaling_analysis}.~~\hyperref[sec:Scaling_analysis]{Complexity and scaling analysis of the method}} \dotfill\textbf{\pageref{sec:Scaling_analysis}}
\medskip

\noindent \textbf{\ref{sec:Noise_analysis}.~~\hyperref[sec:Noise_analysis]{Effects of noise and random quantum circuits}} \dotfill\textbf{\pageref{sec:Noise_analysis}}
\medskip

\noindent \textbf{\ref{sec:metric-MDS}.~~\hyperref[sec:metric-MDS]{Manifold learning of the toric code based on the metric-multidimensional scaling (metric-MDS)}} \dotfill\textbf{\pageref{sec:metric-MDS}}
\medskip

\section{Proof of Theorems}
\label{sec:ProofThms}

\textit{Proof of Theorem~1:} 
For a pure quantum state $\rho(s) = | \psi(s) \rangle \langle \psi(s) |$ in the unitary evolution defined in the QPP of the main text, the QFI for the parameter $s$ reads 
\begin{eqnarray}
\label{eq:QFI}
{\cal{F}}_Q (s) &=& 4 \left( \langle \partial_s \psi(s) | \partial_s \psi(s) \rangle - \left| \langle \partial_s \psi(s) | \psi(s) \rangle \right|^2 \right) \nonumber \\
&=& 4 \left\{ \mathrm{tr} \left( G^2 (s) \rho(s) \right) - \left[ \mathrm{tr} \left( G(s) \rho(s) \right) \right]^2 \right\} \nonumber \\
&=& 4 \text{Var} \left[ G(s)\right],
\end{eqnarray}
where we have used the relation $| \psi(s) \rangle = U(s) | \psi(0) \rangle$ and the hermitian generator $G(s) = i\left[ \partial_s U(s) \right] U^{\dagger}(s)$; and $\text{Var}$ denotes the variance with respect to the \textit{evolved} quantum state $\rho(s)$. The variance of the generator $G(s)$ is upper bounded by  
\begin{eqnarray}
\label{eq:Var_G}
\text{Var} \left[ G(s)\right] \le \frac{1}{4} \left[\lambda_{\text{max}}(s) - \lambda_{\text{min}}(s) \right]^2 \le \| G(s) \|^2_{\infty},
\end{eqnarray}
where $\lambda (s)$ is the eigenvalue of $G(s)$; and to obtain the last inequality, we have used $|a - b| \le |a| + |b|$ for real numbers $a$ and $b$, and that, by definition, the operator norm of the generator $\| G(s) \|_{\infty} = | \lambda (s)|_{\text{max}}$. The first inequality in (\ref{eq:Var_G}) is saturated when $|\psi(s)\rangle = \frac{1}{\sqrt{2}} \left(|\lambda_{\text{max}} (s) \rangle + |\lambda_{\text{min}} (s) \rangle \right)$, i.e., the quantum state is an equal weight superposition of eigenstates with maximal and minimum eigenvalues of $G(s)$, respectively; and the last inequality is saturated when $\lambda_{\text{max}}(s) = - \lambda_{\text{min}} (s) > 0$.

\medskip
On the other hand, the operator norm of the $G(s) = \sum_{\sigma} h_{\sigma} (s) \ \sigma$ of Nielsen's circuit complexity is upper bounded by 
\begin{eqnarray}
\| G(s) \|_{\infty} \le \sum_{\sigma} | h_{\sigma} (s) |,
\end{eqnarray}
where we have used the subadditivity of the operator norm and $\| \sigma \|_{\infty} = 1$ for $n$-qubit Pauli operators $\sigma$. Therefore, we have 
\begin{eqnarray}
\label{eq:FQ_and_Sum_h}
\sqrt{\mathcal{F}_Q(s)} \le 2 \sum_{\sigma} | h_{\sigma} (s) | 
\end{eqnarray}
for \textit{all possible paths} $h_{\sigma} (s)$ ($s \in [0, 1]$). Then by integrating over $s \in [0, 1]$ on both sides of (\ref{eq:FQ_and_Sum_h}) followed by taking the infimum over all possible paths, the first inequality in (8) of the main text holds naturally, i.e., 
\begin{eqnarray}
\mathcal{C}_{\mathcal{F}} \left( \rho_0 \rightarrow \rho_1\right) \le {\cal{C}}_{\mathcal{N}} \left( \rho_0 \rightarrow \rho_1\right).
\end{eqnarray}
The second inequality in (8) of the main text, 
\begin{eqnarray}
\label{eq:C_F_and_Bures}
\frac{D_B \left( \rho_0, \rho_1\right)}{\sqrt{2}} &\le& {\cal{C}}_{\mathcal{F}} \left( \rho_0 \rightarrow \rho_1\right),
\end{eqnarray}
which lower bounds the QFC by the Bures distance between $\rho_0$ and $\rho_1$, can be readily and easily derived from the result in~\cite{TaddeiPRL2013}. 

\textit{The above proof can be directly generalized to systems other than qubits.} Suppose that the generator of the unitary quantum path planning is 
\begin{eqnarray}
G(s) = \sum_{i} h_i (s) \ O_i, 
\end{eqnarray}
where $O_i$ is an orthonormal operator basis with operator norm $\| O_i \|_{\infty} \le 1$, then, by replacing the summation in the right-hand side of (\ref{eq:FQ_and_Sum_h}) to go over all orthonormal operator bases of the generator, 
(\ref{eq:FQ_and_Sum_h}) also holds, i.e., 
\begin{eqnarray}
\label{eq:FQ_and_Sum_h_general}
\sqrt{\mathcal{F}_Q(s)} \le 2 \sum_{i} | h_{i} (s) |. 
\end{eqnarray}

\bigskip
Next, a suitable similarity measure for topological quantum order imposes the additional constraint of geometric locality of the QPP, which requires that $O_i$ is a geometrically local operator for all possible values of $i$, i.e., $O_i$ is supported on a constant number of neighboring particles. We consider a reduced density matrix $\rho \left( s | \Delta \right)$ supported on a subsystem $\Delta$ of constant size, undergoing a nonunitary evolution along the unitary QPP of the total system generated by $G(s)$. While the relation between the Bures distance and the QFI in (6) of the main text still holds for mixed quantum states with the Uhlmann-Jozsa fidelity, the expression of the QFI for $\rho \left( s | \Delta \right)$ with respect to $s$ is more complicated than (\ref{eq:QFI}). Now the QFI has the general form given by the symmetric logarithmic derivative (SLD) $\mathcal{L}$, with 
\begin{eqnarray}
\mathcal{F}_Q \left( s | \Delta \right) = \mathrm{tr} \left[\rho \left( s | \Delta \right) \mathcal{L}^2\right],
\end{eqnarray}
where $\mathcal{L}$ is determined by 
\begin{eqnarray}
\partial_s \rho \left( s | \Delta \right) = \frac{1}{2} \left[\rho \left( s | \Delta \right) \mathcal{L} + \mathcal{L} \rho \left( s | \Delta \right) \right].
\end{eqnarray}

We make a Trotter decomposition of the unitary evolution generated by $G(s)$ into a product of many sufficiently small, finite time steps. At each Trotter time step, we can estimate an upper bound for $\mathcal{F}_Q \left( s | \Delta \right)$ of the subsystem. The evolution of the reduced density matrix of the subsystem $\Delta$ results from a partial-trace operation (with respect to the rest of the system) on the total system which undergoes the unitary evolution generated by $G(s)$. The QFI of the reduced density matrix with respect to $s$ does not exceed that of its enlarged pure system, i.e., 
\begin{eqnarray}
\label{eq:F_Q_purification}
\mathcal{F}_Q \left( s | \Delta \right) \le \mathcal{F}_Q \left( s \right),
\end{eqnarray}
where the latter can be estimated via (\ref{eq:FQ_and_Sum_h_general}) through unitary evolution. Note that (\ref{eq:F_Q_purification}) can be readily obtained with the monotonicity of the Uhlmann-Jozsa fidelity of two quantum states $\rho$ and $\tilde{\rho}$~\cite{NielsenChuang}, i.e.,  
\begin{eqnarray}
F\left[\mathcal{E}(\rho), \ \mathcal{E}(\tilde{\rho}) \right] \ge F\left(\rho, \ \tilde{\rho} \right)
\end{eqnarray}
under the trace-preserving quantum operation $\mathcal{E}$, which means that the Bures distance and QFI are contractive under the partial-trace operation.

Due to the geometric locality of the generator and of the quantum circuit applied to the system as well as the \textit{very small evolved time step}, the generators $\{O_i \}$ that govern the time evolution of the subsystem $\Delta$ are restricted to include only those supported on $\Delta$ and its neighboring environment, where the latter is directly coupled to $\Delta$ by applying the quantum circuit. We denote $\tilde{\Delta}$ as the subsystem $\Delta$ and its neighboring environment, where the size $| \tilde{\Delta} | \le \text{constant}$. Consequently, the summation on the right-hand side of (\ref{eq:FQ_and_Sum_h_general}) for upper bounding $\mathcal{F}_Q \left( s | \Delta \right)$ must be restricted to the basis operators which are locally supported on $\Delta$ and its neighboring environment, i.e., 
\begin{eqnarray}
\frac{1}{2} \sqrt{ \mathcal{F}_Q \left( s | \Delta \right) } \le \frac{1}{2} \sqrt{\mathcal{F}_Q \left( s \right)} 
\le \underset{i \ : \ \mathcal{S}_i \subset \tilde{\Delta} }{\sum} | h_i(s)|,
\end{eqnarray}
where $\mathcal{S}_i$ is the support (on the system) of the $i$-th orthonormal basis operator $O_i$ in the expansion of the generator $G(s)$. 

Then, by integrating over $s \in [0, \, 1]$, followed by summing over different such subsystems of constant size to cover the whole system in order to cover the support of the generator, we have 
\begin{eqnarray}
\label{eq:bound_QFI}
\underset{\Delta}{\sum} \ \frac{1}{2} \int_0^1 \sqrt{ \mathcal{F}_Q \left( s | \Delta \right) } \ \mathrm{d}s
\le  \int_0^1 \ \underset{\Delta}{\sum} \ \ \underset{i \ : \ \mathcal{S}_i \subset \tilde{\Delta} }{\sum} | h_i(s)| \  \mathrm{d}s \approx \int_0^1 \underset{i}{\sum} | h_i(s)| \ \mathrm{d}s,
\end{eqnarray}
where the summation in the right-hand side of the last approximate equation goes over all operator bases of the generator. Moreover, from the result presented in~\cite{TaddeiPRL2013}, we can derive that 
\begin{eqnarray}
\frac{1}{\sqrt{2}} D_B \left[ \rho_0 \left( \Delta \right), \rho_1 \left( \Delta \right) \right] \le \frac{1}{2} \int_0^1 \sqrt{ \mathcal{F}_Q \left( s | \Delta \right) } \ \mathrm{d}s,
\end{eqnarray}
which always holds for arbitrary smooth paths between the two reduced density matrices, caused by the smooth unitary path of the total system followed by the partial-trace operation, i.e., it is path-independent. Then we can take the infimum over all possible paths of the unitary QPP on the right-hand side of the last equation in (\ref{eq:bound_QFI}) to obtain a tighter bound, and conclude that Nielsen's QCC of geometrically local quantum circuits is approximately lower bounded by the summation over fidelity distances of the reduced density matrices supported on non-overlapping subsystems $\{\Delta \}$ which, together with their neighboring environments, cover the total system, i.e., 
\begin{eqnarray}
\mathcal{C}_{\mathcal{N}} \left( \rho_0 \rightarrow \rho_1\right) \gtrsim \frac{1}{\sqrt{2}} \, \underset{\Delta}{\sum} \, D_B \left[ \rho_0 \left( \Delta \right), \rho_1 \left( \Delta \right) \right].
\end{eqnarray}

(This completes the proof of Theorem~1.) 

\bigskip

\textit{Proof of Theorem ~2:} 
Here we follow a similar procedure as in~\cite{EisertPRL2021}, but with more details in order to generalize the result to two arbitrary multi-qubit states in a generic QPP problem.  Arrange the $n$-qubit system into a qubit chain $\{1, 2, \cdots, n \}$ following a specific order, and take a subsystem over the $k$-th cut, $A_k = \{1, 2, \cdots, k \}$, and its environment $B_k = \{k+1, k+2, \cdots, n \}$ as in the statement before Theorem~2. Further,  we assume that the coupling in the generator $G(s)$ of $U(s)$ between $A_k$ and $B_k$ is spatially local with respect to the qubit chain. By Trotter decomposition, slicing $s \in [0, \, 1]$ into small time steps $\{ s_j \}_{j=0}^{T_s - 1}$, and at each time interval $\Delta s_j = \left|s_{j} - s_{j-1}\right|$ ($j \in [1, T_s - 1]$), the unitary evolution operator of the QPP is given by
\begin{eqnarray}
U(s_j) = \mathrm{exp} \left[-i \Delta s_j G(s_j) \right] .
\end{eqnarray}
The generators $G(s_j)$ can be decomposed as 
\begin{eqnarray}
G(s_j) = G_{A_k}(s_j) + G_{B_k}(s_j) + G_{k}(s_j),
\end{eqnarray}
where $G_{A_k}(s_j)$  and $G_{B_k}(s_j)$ are supported on $A_k$ and $B_k$, respectively, and 
\begin{eqnarray}
G_k(s_j) = \underset{O^{(k)}}{\sum} h_{O^{(k)}}\left(s_j\right) O^{(k)},
\end{eqnarray}
is the local coupling between the subsystems $A_k$ and $B_k$, with $O^{(k)}$ a geometrically local basis operator at the $k$-th bond of operator norm $\left\| O^{(k)}\right\|_{\infty} \le 1$. With rigorous results already proved in~\cite{BravyiPRL2006,BravyiPRA2007,HutterPRL2012,Van_AcoleyenPRL2013,MariCMP2016} concerning the \emph{Small Incremental Entangling (SIE)}, the rate of the entanglement over the cut $k$ generated or destroyed in this small time slice is upper bounded by 
\begin{eqnarray}
\label{eq:SIE}
\left|\frac{\mathrm{d}}{\mathrm{ds}} S_k \left[\rho(s) \right] \right| \, \Bigg|_{s = s_j} 
&\le& c^{\prime} \, \ln d_k \, \left\| G_k (s_j) \right\|_{\infty}  \nonumber \\
&\le& c^{\prime} \, \ln d_k \, \underset{O^{(k)}}{\sum} \left| h_{O^{(k)}}\left(s_j\right) \right|,
\end{eqnarray}
where $c^{\prime} > 0$ is a constant and $d_k = \min \{d_{A_k}, d_{B_k}\} \le 2^{n-1}$ is the minimal Hilbert-space dimension of the subsystems. With (\ref{eq:SIE}), integrating over $s \in [0, \, 1]$ (summing over all infinitestmal time slices $\Delta s_j$) and using the relation $\left| \int_0^1 f(s) \mathrm{d}s\right| \le \int_0^1 \left| f(s) \right| \mathrm{d}s$ for smooth functions, followed by summing over all cuts $k$, we obtain that Nielsen's circuit complexity measure is, for all possible paths, lower bounded by the entanglement change averaged over the bonds. We can then take the infimum of the circuit complexity measure over all paths allowed to obtain a tighter bound, i.e., 
\begin{equation}
\frac{c}{\left(n - 1 \right)} \sum^{n-1}_{k=1} \left| S_k (\rho_1) - S_k (\rho_0) \right| \le \mathcal{C}_{\mathcal{N}} \left( \rho_0 \rightarrow \rho_1\right),
\end{equation}
for some constant $c = 1/\left( c^{\prime} \ln 2\right) > 0$.
Note that here the reference quantum state $\rho_0$ is not restricted to a product state, and we find an averaging prefactor $1/(n-1)$ compared to the result in~\cite{EisertPRL2021}. 
(This completes the proof of Theorem~2.)

\section{Topological equivalence of gapped ground states and quantum circuit complexity of adiabatic quantum computation}
\label{sec:AQC}
As mentioned in Definition~2, two gapped quantum states $\rho_0$ and $\rho_1$ in the same topological phase can be connected via a smooth path $\rho(s) = U(s) \rho_0 U^{\dagger}(s)$ for $s \in [0, 1]$. In case of gapped spectra, an explicit expression for the generator $G(s)$ is provided in~\cite{BachmannCMP2011}, within the spectral flow formalism:
\begin{eqnarray}
\label{eq:GappedGenerator}
G(s) &=& \int_{-\infty}^{+\infty} \mathrm{d}t \ W_{\gamma}(t) \ \mathrm{e}^{it H(s)} 
\, \partial_s H(s) \, \mathrm{e}^{-it H(s)},
\end{eqnarray}
where $H(s)$ is the Hamiltonian path with a spectral gap $\delta(s) \ge \gamma > 0 \left(\forall s \in [0, 1]\right)$ and with the ground state $\rho(s)$; $W_{\gamma}(t) \in L^1 (\mathbb{R})$ is a decaying weight function of the spectral flow satisfying $\left\| W_{\gamma} \right\|_1 \le c_0/\gamma$, for some positive constant $c_0$. In this situation, the Nielsen QCC may be explicitly obtained by optimization based on differential geometry, which deserves further investigation.

From the Proof of Theorem~1, we know that the integrand of Nielsen's QCC is lower bounded by the operator norm of the generator, with $\| G(s) \|_{\infty} \le \sum_{\sigma} | h_{\sigma} (s) |$. This inequality is already sufficiently satisfied if we replace $\| G(s) \|_{\infty}$ on the left-hand side by its maximal value. With (\ref{eq:GappedGenerator}) and the properties of the decaying weight function of the spectral flow, one finds the maximal value to satisfy
\begin{eqnarray}
\label{eq:UpperBoundG}
\| G(s) \|_{\infty} \le \frac{c_0 \| \partial_s H(s) \|_{\infty}}{\gamma}. 
\end{eqnarray}
\mdf{
It follows that Nielsen's QCC can be estimated as 
\begin{eqnarray}
\label{eq:QCC_Order}
\mathcal{C}_{\mathcal{N}} \sim \mathcal{O} \left( \underset{s \in [0, 1]}{\mathrm{max}} \frac{\| \partial_s H(s) \|_{\infty}}{\gamma} \right), 
\end{eqnarray}
which is similar (but not identical) to the adiabaticity condition in adiabatic quantum computation~\cite{AQCRMP2018}.

\section{Asymptotic and tightness analysis of the bounds in Theorems 1 and 2}
\label{sec:Tightness_analysis}
For gapped ground states, we will show in the following that the quantum Fisher complexity (QFC) in Theorem 1 is asymptotically (with respect to the number of qubits $n$) tight as a lower bound of the Nielsen's QCC. For the quantum path planning following the gapped ground state $\rho(s)$ of the Hamiltonian $H(s)$, one can write the quantum Fisher information of the ground state $\rho(s) = |\psi_0(s)\rangle \langle \psi_0(s)| $ as~\cite{YouPRE2007,GietkaQuantum2021}
\begin{eqnarray}
\label{eq:QFI_adiabatic}
{\cal{F}}_Q (s) &=& 4\sum_{m\neq 0}\frac{|\langle \psi_m(s)|\partial_s H(s)|\psi_0(s)\rangle|^2}{[E_m(s)-E_0(s)]^2},
\end{eqnarray}
where the Hamiltonian has a spectral representation $H(s) = \sum_{m=0} E_m(s) |\psi_m(s)\rangle \langle \psi_m(s)|$ with $E_m(s) \le E_{m+1}(s) $ for $m=0, 1, 2, \cdots$, and a spectral gap $E_1 (s) - E_0 (s) \ge \gamma > 0 \left(\forall s \in [0, 1]\right)$. Note that the spectral gap $\gamma$ can scale as $1/\mathrm{poly}(n)$ or $1/\mathrm{exp}(n)$. 

With (\ref{eq:QFI_adiabatic}), an upper bound for the quantum Fisher information can be readily obtained, 
\begin{equation*}
\begin{split}
{\cal{F}}_Q(s) &\leq \frac{4}{\gamma^2} \sum_{m \neq 0} \langle \psi_0(s) | \partial_s H(s) | \psi_m(s) \rangle \langle \psi_m(s) | \partial_s H(s) | \psi_0(s) \rangle \\
&= \frac{4}{\gamma^2} \langle \psi_0(s) | \partial_s H(s) \left( \mathcal{I} - |\psi_0(s)\rangle \langle \psi_0(s)| \right) \partial_s H(s) | \psi_0(s) \rangle \\
&= \frac{4}{\gamma^2} \text{Var}_{|\psi_0(s)\rangle} [\partial_s H(s)] \\
&\leq \frac{4}{\gamma^2} \| \partial_s H(s) \|_\infty^2,
\end{split}
\end{equation*}
where $\mathcal{I} = \sum_{m=0} |\psi_m(s)\rangle \langle \psi_m(s)|$ is the identity, and we have used (\ref{eq:Var_G}) to obtain the last inequality.

Therefore, the QFC is 
\begin{eqnarray}
\label{eq:QFC_Order}
\mathcal{C}_{\mathcal{F}} \sim \mathcal{O} \left( \underset{s \in [0, 1]}{\mathrm{max}} \frac{\| \partial_s H(s) \|_{\infty}}{\gamma} \right), 
\end{eqnarray}
which is asymptotically same as Nielsen's QCC $\mathcal{C}_{\mathcal{N}}$ in (\ref{eq:QCC_Order}). The scaling with respect to the system size is dominated by the spectral gap $\gamma \sim 1/\mathrm{poly}(n)$ or $\gamma \sim 1/\mathrm{exp}(n)$. 

We have shown that the QFC is asymptotically tight to the Nielsen's QCC, while the Bures distance (or the fidelity) is actually less stringent, because the Bures distance between two quantum states has a constant upper bound. To study topological phases, we need to further incorporate the condition of geometric locality, which leads to the inequality (9) in Theorem 1 of the main text, where the lower bound of $\mathcal{C}_{\mathcal{N}}$ is asymptotically $\mathcal{C}_{\mathcal{N}} \ge \Omega (n)$. On the other hand, for the entanglement-based lower bound of $\mathcal{C}_{\mathcal{N}}$ in Theorem 2, we have 
\begin{eqnarray}
\label{eq:Omega_Sk}
\mathcal{C}_{\mathcal{N}} \ge \Omega \left( \underset{k \in [1, n-1]}{\mathrm{max}} \left| S_k (\rho_1) - S_k (\rho_0) \right| \right). 
\end{eqnarray}
Therefore, for generating the (two-dimensional) Kitaev's toric code (which has area-law quantum entanglement) from a product state, this lower bound is $\mathcal{O} \left(L\right)$, where $L$ is the system size. Since the known optimal QCC for the toric code is $\mathcal{O} \left(L\right)$~\cite{BravyiPRL2006}, our entanglement-based bound is asymptotically tight for the toric code ground state, while the fidelity-based one overestimates the quantum circuit cost. On the other hand, for quantum states with volume-law quantum entanglement, we obtain similar lower bounds as in Theorem 1, that is, $\mathcal{O} \left(n\right)$. The (gapped) ground state of the bond-alternating XXZ spin chain in the main text has short-range quantum entanglement, whose quantum circuit depth / complexity is a constant, i.e., not scaling with $n$. The Von Neumann entropy (area-law scaling) of this ground state is also a constant. In this case, both the fidelity lower bound in (9) of Theorem 1 (after normalized by $n$ in the kernel) and the entanglement lower bound in (\ref{eq:Omega_Sk}) are asymptotically tight to the Nielsen's QCC. For more exotic and complicated quantum states, whose Nielsen's QCC may grow exponentially with $n$, both lower bounds [see Eq. (9) in Theorem 1 and Eq. (12) in Theorem 2] are no longer tight in this case. More elaborations will be made in future work to address such quantum states, as well as the information losses in the approximations of Nielsen's QCC. 

\section{Complexity and scaling analysis of the method}
\label{sec:Scaling_analysis}
As we have discussed in Sec.~\ref{sec:Tightness_analysis}, the quantum Fisher complexity is asymptotically tight to Nielsen's QCC, and contains information about topological properties, because the quantum Fisher information (QFI) reflects entanglement depth and critical behavior of the quantum state during the quantum path planning. In contrast, the Bures distance (and fidelity) is less stringent but more efficient in measurement and computation. For instance, in the numerical demonstrations of the main text, we use two-body reduced density matrices of constant size to cover the whole system, for which the computational or the measurement complexity in calculating the kernel is $\mathcal{O} (n)$.  
On the other hand, the entanglement-based kernel derived from Theorem 2 contains more topological information compared to the fidelity-based one. This is consistent with the knowledge that QFI and entanglement entropy are related to global properties of quantum states, which are \textit{generically} hard to access through classical computation or measurement. For instance, one prominent approach based on classical measurement of quantum-state properties is the classical shadow tomography~\cite{HuangNatPhys2020}, where nonlinear functional of the density matrix can be estimated from randomized Pauli measurement (classical shadows). In particular, QFI and Von Neumann entropy can be approximated~\cite{RathPRL2021,  VermerschPRXQ2024}, respectively, as a series expansion of polynomials of the density matrix, up to a prescribed precision. This involves estimations of high-order moments of the density matrix, which in general require exponentially many copies of the quantum states. Reference~\cite{VermerschPRX2024} shows that polynomially many local measurements can be sufficient for estimating the second-order moment (or the R\'enyi entropy) if the quantum state posesses certain structures and properties, while it remains open for the Von Neumann entropy and the QFI, which deserves further investigations.

Therefore, we may conclude that topological order is hard to learn in the worst case. Alternatively, for those quantum states that are efficiently simulatable via tensor networks or matrix product states, the entanglement-entropy profile (which is related to the bond dimensions) is automatically stored and updated during the ground-state search process of the DMRG algorithm. This makes the entanglement-based kernel accessible efficiently. This point is also mentioned in the main text. 

On the other hand, in case that our input data for the kernel are quantum states prepared on quantum devices, quantum algorithms may help to reduce the scaling and complexity of calculating the entanglement-based kernel. For instance, the quantum algorithm based on a Fourier series of the density matrix proposed in~\cite{WangPRAppl2023} can estimate the Von Neumann entropy with $\mathcal{O} \left[ \mathrm{poly} \left( \frac{1}{\epsilon}, \ \frac{1}{\Lambda} \right) \right]$ copies of the quantum state and $\mathcal{O} \left[ \mathrm{poly} \left( n, \frac{1}{\epsilon}, \ \frac{1}{\Lambda} \right) \right]$  single- / two-qubit quantum gates, where $\epsilon$ is the approximation precision and $\Lambda$ is the lower bound of non-zero eigenvalues of the density matrix. The algorithm provides a quantum speedup in case that $\Lambda$ is a constant or is $\Omega \left[ 1 / \mathrm{poly} \left( n \right) \right]$; Alternatively, the density-matrix-exponentiation (DME) algorithm~\cite{LloydNatPhys2014} can help to estimate the entanglement entropy of low-rank quantum states efficiently in the framework of Hamiltonian learning; etc. 
These results bring promise for quantum speedups in estimating the entanglement-based kernel within a quantum-classical hybrid learning framework. Moreover, estimating nonlinear functionals of the density matrix such as the QFI and the entanglement entropy efficiently on either classical, quantum, or hybrid platforms is still an active research area of quantum information and computation. 

Finally, suppose that we have the number of samples (quantum states) $N$ and a $N \times N$ kernel matrix, the (classical) computational complexity of the kernel PCA and the diffusion map (manifold learning) algorithm is $\mathcal{O} \left( N^3 \right)$, which involves spectral analysis of the kernel matrix. This can be efficient for small data sets. For large data sets (e.g., $N > 100$), either the metric-MDS or the t-SNE for manifold learning is more suitable, the computational complexity of which is $\mathcal{O} (T N^2)$, where $T$ is the number of iterations in the learning algorithm. (This completes the scaling analysis of our machine learning method.)

\begin{figure*}[t]
\centerline{\includegraphics[height=3.2in,width=7in,clip]{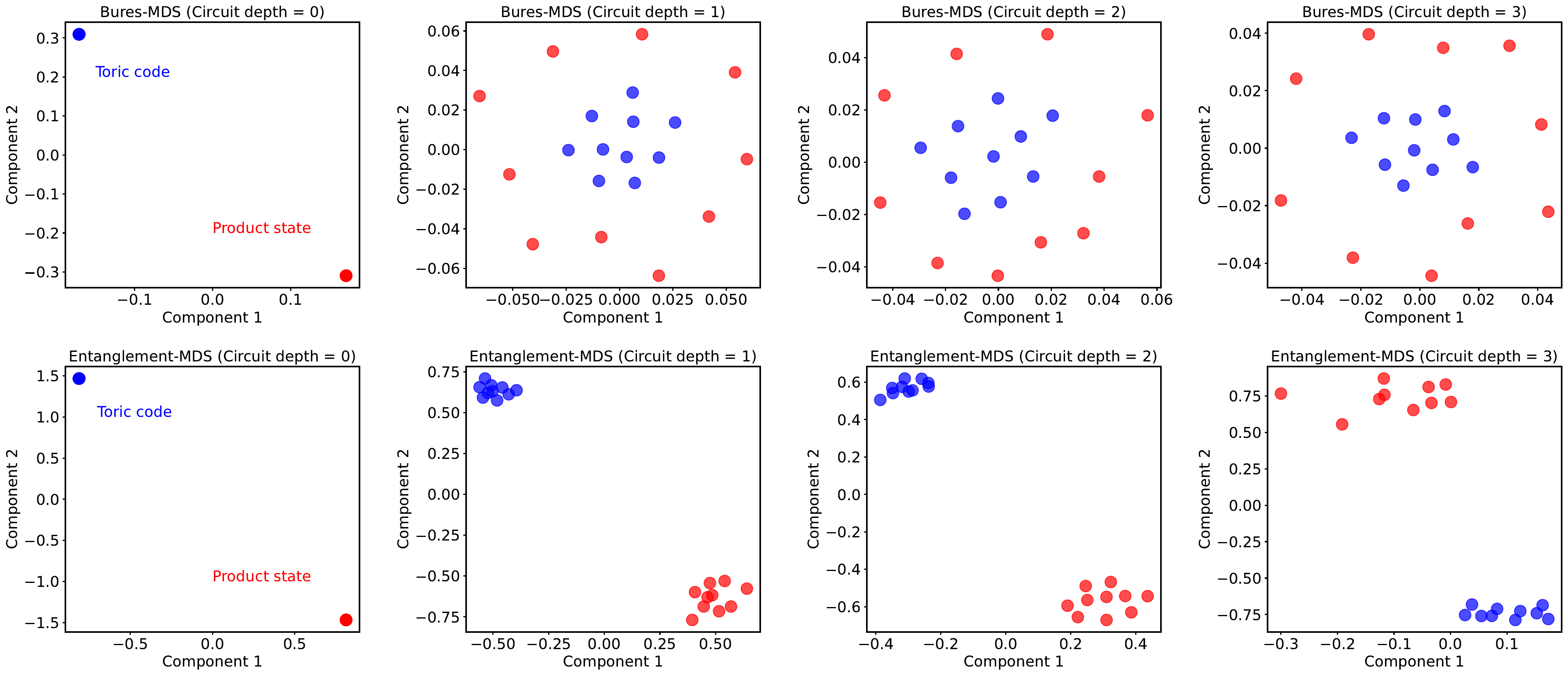}}
\caption{\textbf{Unsupervised manifold clustering of the ground state of Kitaev's toric code (blue dots in all sub-figures) and product states (red dots in all sub-figures) via the Bures distance (four sub-figures on the top) and entanglement distance (four sub-figures in the bottom) based metric-MDS, respectively.} Two-qubit Haar random gates~\cite{tenpy} of different circuit depth (Circuit depth $= 0, 1, 2, 3$) are applied repeatedly to both the toric code and the product state $| 0\rangle ^{\otimes n}$. We use $N = 20$ samples in total, with $10$ samples of the toric code (blue dots) and $10$ samples of the product state (red dots). We use $n = 32$ qubits (on a lattice of size $4 \times 4$ with toric boundary conditions; i.e., with a code distance of $4$), and the DMRG (with the SVD cutoff $10^{-10}$ and the maximal energy error $10^{-10}$) to solve for the ground state of the toric code and, at the same time, to extract the entanglement profile. We use $n$ geometrically local $2$-body reduced density matrices to cover the system and to calculate the local Bures distance in Eq. (9) of Theorem 1 in the main text. The result shows that, in the absence of noise (i.e. with circuit depth zero), both the Bures- and the entanglement-metrics perform well in clustering the toric code and the product states, while the entanglement-MDS is more robust to Haar random noises, indicating that the entanglement metric is a better choice in both the clustering performance and the interpretability.
}
\label{fig:TC_metric-MDS}
\end{figure*}

\section{Effects of noise and random quantum circuits}
\label{sec:Noise_analysis}
Regarding potential effects of hardware noise, one may basically consider three scenarios. 

The first scenario is that decoherence such as dephasing occurs during the quantum path planning or the quantum state preparation. For instance, the quantum Fisher information (i.e., the fidelity susceptibility) can scale as $\mathcal{O} \left( \mathrm{poly} (n) \ \mathrm{e}^{-\gamma n}\right)$ (under independent dephasing) or scale as $\mathcal{O} \left( \mathrm{poly} (n) \ \mathrm{e}^{-\gamma n^2}\right)$ (under correlated collective dephasing) for certain state evolution paths~\cite{ChePRA2019}, where $\gamma$ is the noise strength. This may expain why the fidelity-based kernel is more fragile to noises. Generically, the presence of noise invalidates our original ansatz of unitary path planning, where the optimal distance is defined on a unitary-group manifold. Proper topological distance between noisy quantum states may require the analysis of the complexity of local quantum channels, a generalization of our current approach, which we leave for future work. 

The second scenario is that, once the topologically ordered quantum state (e.g., Kitaev's toric code) is prepared, for instance, under quantum error mitigation or error correction, the established long-range entanglement will be robust to local decoherence. The Haar random noises uniformly apply random rotations or couplings to qubits, which is usually used for simulating and performance benchmarking of quantum information processing under strong hardware noise levels (e.g., depolarization). In this respect, the robustness against the Haar random two-qubit gates applied to the toric code in the main text (see Fig. 3 in the main text) indicates the superior performance of the entanglement-based kernel. 

Finally, for an ensemble of random quantum states $\rho$ (e.g., generated from random quantum circuits with the initial state $\rho_0$) described by a parametrized  distribution $p_\theta$, the ensemble-averaged kernel of the Nielsen's QCC is given by 
\begin{eqnarray}
\langle \ \mathcal{K_{\mathcal{N}}} \ \rangle_{p_{\theta}} = \int \ \mathrm{e}^{-\beta \ \mathcal{C}_{\mathcal{N}} \left(\rho_0, \ \rho \right)} \ p_{\theta} \left(\rho\right) \mathrm{d} \rho,
\end{eqnarray}
which plays a role similar to the generating function~\cite{Miyajiarxiv2025} of the (averaged) quantum circuit complexity, 
\begin{eqnarray}
\langle \ \mathcal{C}_{\mathcal{N}} \ \rangle_{p_{\theta}} = - \lim_{\beta \to 0} \ \frac{\partial}{\partial \beta} \langle \ \mathcal{K_{\mathcal{N}}} \ \rangle_{p_{\theta}}.
\end{eqnarray}
This is similar to the action-propagator relation in generative modelling of Feynman paths via deep learning approach~\cite{ChePRB2022}, and can be useful in optimizing variational quantum circuits in various quantum machine learning tasks. 

\section{Manifold learning of the toric code based on the metric-multidimensional scaling (metric-MDS)}
\label{sec:metric-MDS}

Diffusion maps and kernel PCA that we have used in the main text involve a kernelization based on the distance metric, where the kernel is used to construct a conditional probability in the diffusion map (as well as in the t-SNE algorithm) and is directly diagonalized in the kernel PCA. While this kernelization may complicate the understanding of the distance metrics, we can use an alternative manifold learning called the metric-multidimensional scaling (metric-MDS)~\cite{DemainePMLR2021,SklearnMDS}, where high-dimensional manifold of the data set with the Bures or entanglement distance metric can be embedded into a low dimensional Euclidean space, while preserving the original distance relations of the data manifold \textit{without involving a kernel}. This can be realized, for instance, by minimizing a stress function (cost function) iteratively~\cite{DemainePMLR2021,SklearnMDS}. \textit{The metric-MDS does not require a kernel, but utilizes the distance metric of the manifold directly as in the Isomap~\cite{TenenbaumScience2000}.} In Fig.~\ref{fig:TC_metric-MDS}, we plot the comparison of the Bures-MDS and the entanglement-MDS manifold learning for the toric code. We find that, in the absence of noise (i.e. with circuit depth zero), both the Bures- and the entanglement-metrics perform well in clustering the toric code and the product states. After applying two-qubit Haar random local quantum gates, the entanglement-MDS is more robust in the clustering performance, indicating that the entanglement metric is a better choice accounting for both the clustering performance and the interpretability (i.e., it contains more topological information). More details can be found in the caption of Fig.~\ref{fig:TC_metric-MDS}.

}

\bibliography{Ref}

\begin{thebibliography}{125}
\bibitem [{Ily()}]{IlyaNote}%
  \BibitemOpen
  \href@noop {} {}\bibinfo {note} {The idea that the conditional Kolmogorov
  complexity $K(x|y)$ provides a theoretically optimal solution to unsupervised
  machine learning was, to the best of our knowledge, first formulated in a
  talk by Ilya Sutskever.}\BibitemShut {Stop}%
\bibitem [{Not()}]{Note_RDM}%
  \BibitemOpen
  \href@noop {} {}\bibinfo {note} {In order to connect to the classical shadow
  kernel learning, here we do not incorporate the constraint of geometric
  locality for the two-body reduced density matrices used in calculating the
  fidelity-based kernel. Moreover, with geometrical locality, we find that the
  clustering performance of the fidelity-based kernel PCA will be slightly
  boosted, but is still inferior to that of the entanglement-profile based one
  under Haar random noises.}\BibitemShut {Stop}%
\end{thebibliography}

\end{document}